\documentclass[twocolumn,english,aps,prl,twocolumn,showpacs,superscriptaddress,groupaddress,floatfix,graphics,graphicx]{revtex4-1}
\usepackage[T1]{fontenc}
\usepackage[latin9]{inputenc}
\usepackage{color}
\usepackage{babel}
\usepackage{amsmath}
\usepackage{amssymb}
\usepackage{graphicx}
\usepackage{wasysym}
\usepackage[unicode=true,pdfusetitle,
 bookmarks=true,bookmarksnumbered=false,bookmarksopen=false,
 breaklinks=false,pdfborder={0 0 1},backref=false,colorlinks=false]
 {hyperref}
\hypersetup{
 colorlinks,linkcolor=blue,citecolor=blue,urlcolor=blue}
\begin{document}
\title{High-resolution real-space evaluation of the self-energy operator
of disordered lattices: Gade singularity, spin--orbit effects and
p-wave superconductivity}
\author{S. M. João}
\affiliation{Department of Physics, University of York, YO10 5DD, York, United
Kingdom}
\affiliation{Centro de Física das Universidades do Minho e Porto, University of
Porto, 4169-007 Porto, Portugal}
\author{J. M. Viana Parente Lopes}
\affiliation{Centro de Física das Universidades do Minho e Porto, University of
Porto, 4169-007 Porto, Portugal}
\author{Aires Ferreira}
\email{aires.ferreira@york.ac.uk}

\affiliation{Department of Physics and York Centre for Quantum Technologies, University
of York, YO10 5DD, York, United Kingdom}
\begin{abstract}
Disorder is a key factor influencing the behavior of condensed states
of matter, however the true extent of its impact is generally difficult
to determine due to the prominent roles played by quantum interference,
entanglement between spin and orbital degrees of freedom and proximity
to quantum critical points. Here we show that the one-particle disorder
self-energy --- a direct probe of the renormalization of low-energy
excitations due to defects and impurities distributed randomly in
a crystal --- can be obtained by means of unbiased spectral expansions
of lattice Green's functions in a computationally expedient manner.
Our scheme provides a powerful framework to map out the frequency
and wavevector dependence of electronic excitations in unprecedented
large tight-binding systems, up to $10^{9}$ orbitals, with energy
resolution only limited by the mean level spacing. We demonstrate
the versatility of our approach in 3 distinct problems: (i) the Gade
singularity in honeycomb layers with dilute topological defects; (ii)
the rich landscape of impurity resonances in a spin--orbit-coupled
ferromagnet; and (iii) the tailoring of emergent $s$-wave and $p$-wave
superconducting phases in graphene via atomic defects. These examples
reveal rich features in the disorder self-energy $\Sigma_{\textrm{dis}}(\mathbf{k},\omega)$
that are absent from the self-consistent $T$-matrix approach and
other common approximation schemes, which include regimes of nontrivial
wavevector dependence and anomalous dependence upon the impurity concentration.
Our study unravels puzzling, and so far largely inaccessible, manifestations
of strong nonperturbative quantum interference effects in quantum
materials and disordered phases of matter.
\end{abstract}
\maketitle

\section{Introduction}

Disorder plays an essential role in the wealth of phenomena observed
in condensed matter. On the one hand, disorder influences the structural,
optical and transport properties of metals and semiconductors, as
well as the phase stability of unconventional states of matter, such
as chiral supercondutors \citep{Kallin_2016} and symmetry-protected
topological insulators \citep{RevModPhys.83.1057}. On the other hand,
certain types of disorder can induce interesting behavior unseen in
clean systems, ranging from the breakdown of the Fermi liquid description
and quantum interference effects in mesoscopic conductors to the Kondo
effect and novel quantum phases in strongly correlated systems \citep{Sandvik_95,Yu_05,Watanabe_14,Kimchi_18,VARMA2002267,Pires_19}.

Green's functions provide a powerful mathematical device to study
the impact of disorder and correlations in many-body quantum systems.
Of paramount importance in this context is the single-particle irreducible
self-energy which, in effect, dresses bare Green's functions with
the cooperative effects experienced by quasiparticles, and therefore
determines the key features of the spectral function measured in angle-resolved
photoemission experiments \citep{Norman_99,RevModPhys.93.025006}.
The disorder contribution to the self-energy in a homogeneous disordered
system is defined in terms of disorder-free, $\hat{G}_{0}(\mathbf{k},\omega)$,
and full, $\hat{G}(\mathbf{k},\omega)$, single-particle Green's functions
as $\hat{\Sigma}_{\textrm{dis}}(\mathbf{k},\omega)=\hat{G}_{0}^{-1}(\mathbf{k},\omega)-\langle\hat{G}(\mathbf{k},\omega)\rangle^{-1}$
(here $\langle...\rangle$ indicates configurational ensemble average)
and contains detailed information on the effects of quenched disorder,
such as the existence of impurity bound states, as well as the precise
extent of quasiparticle renormalization induced by single impurity
scattering events and quantum interference effects. Because the self-energy
is intimately connected to vertex functions in diagrammatic theory
\citep{Ward_50}, it is also an indispensable tool in the study of
quantum transport phenomena; most notably as a means to obtain conserving
approximations in self-consistent field-theoretical calculations \citep{RevModPhys.46.465}.

The self-energy from electron--impurity interactions, as is presented
in textbooks, is usually described in terms of a complex scalar, $\hat{\Sigma}_{\textrm{dis}}(\mathbf{k},\omega)=\varDelta(\mathbf{k},\omega)\mp i/[2\tau(\mathbf{k},\omega)]$,
where the real part is responsible for the effective-mass renormalization
and $\tau(\mathbf{k},\omega)$ is the quasiparticle elastic scattering
lifetime (here the signs $\mp$ hold for retarded/advanced Green's
functions as required by analyticity). Closed analytical expressions
for the single-particle self-energy can be obtained by means of the
diagrammatic technique based on perturbative expansions in the semiclassical
parameter $g\propto(\omega\tau)^{-1}$\citep{Mahan}. The classical
example is a Fermi gas with a dilute concentration of impurities,
for which a lowest-order ``rainbow'' diagram calculation yields
the relaxation time $\tau(\omega)\propto[n\,u_{0}^{2}\,\nu(\omega)]^{-1}$,
with $n$ the impurity density, $u_{0}$ the potential strength and
$\nu(\omega)$ the bare density of states. The weak-disorder approximation
can be improved systematically by partial resummation of infinite
series of diagrams (e.g., by means of the $T$-matrix or coherent
potential approximations) and extended to multi-orbital Hamiltonians
for both noninteracting and interacting cases; for a recent review
see Ref. \citep{RevModPhys.90.025003}. 

Despite its notable successes, the diagrammatic approach suffers from
a major drawback: the topological complexity of scattering diagrams
increases quickly with the perturbation order, which presents a formidable
barrier to our understanding of nontrivial manifestations of disorder
beyond the weak-coupling regime. Indeed, a range of intriguing phenomena
triggered by strong (nonperturbative) disorder effects are currently
the focus of intense investigation. Some examples include strong Anderson
localization \citep{RevModPhys.80.1355,Karpiuk_12,Gosh14,Marinho18,Richard19},
rare region effects in three-dimensional topological semimetals \citep{Nandkishore14,Pixley16,Pires_21,Pixley21}
and frozen multifractality in chiral-symmmetric lattices \citep{Motrunich_02,Mudry_03,ostrovskyDensityStatesTwoDimensional2014,Hafner_14,Sbierski_20},
whose analysis has defied even the most advanced field-theoretic approaches.
At the heart of such unusual phenomena is the nonperturbative accumulation
of quantum coherent scattering processes, whose satisfactory description
calls for the use of large-scale numerical approaches.

Here, we report a comprehensive numerical study of strong-coupling
effects on the disorder self-energy in several electronic phases of
matter. Our study is based on a new high-resolution real-space spectral
method that gives access to previously unexplored features of the
disorder self-energy, unveiling its rich internal matrix structure
and full $\mathbf{k}$- and $\omega$- dependences. Our calculations,
performed on very large systems, disclose several important features
of the quasiparticle self-energy that are absent in the standard perturbative
treatments, including surprisingly strong $\mathbf{k}$-dependence
generated by nonlocal correlations (quantum interference) and the
emergence of off-diagonal self-energy elements near quantum criticality.

\begin{figure*}
\begin{centering}
\includegraphics[width=1\textwidth]{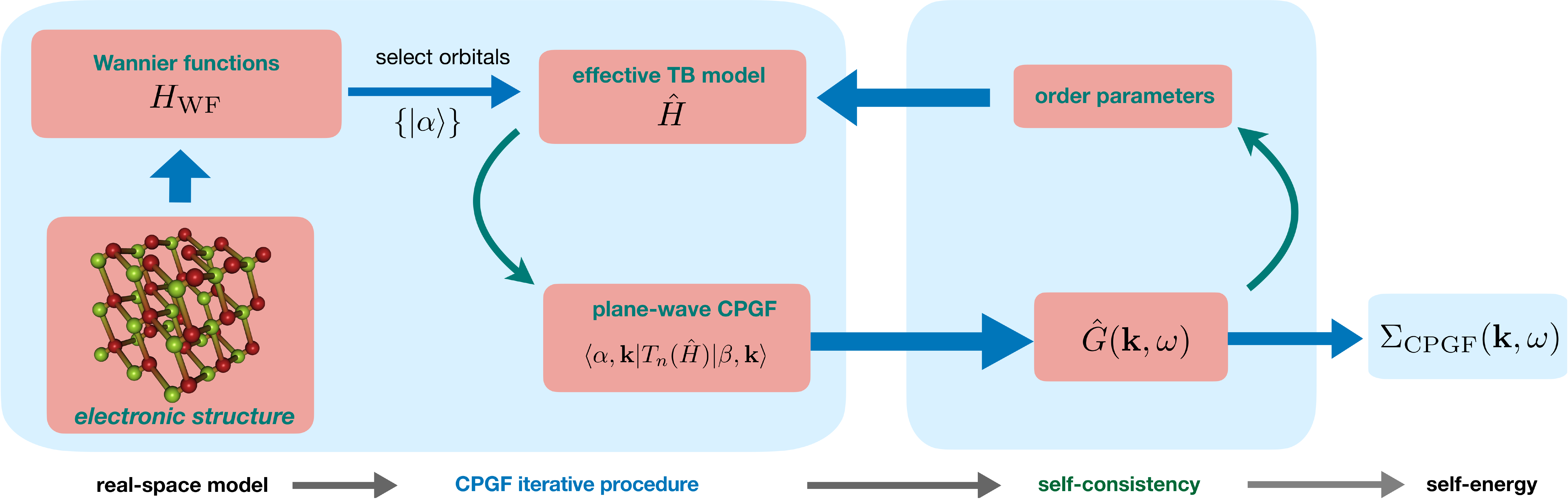}
\par\end{centering}
\caption{\label{fig:01}Bird's-eye view of the spectral evaluation of the $\mathbf{k}$-space
resolved disorder self-energy matrix. The workflow consists of three
steps: (i) derivation of a minimal-basis Hamiltonian that captures
the relevant features of the disordered compound; (ii) large-scale
evaluation of Chebyshev moments of the Hamiltonian matrix; and (iii)
high-resolution spectral reconstruction of self-energy operator using
an exact Chebyshev decomposition of the lattice Green's function.
For problems that require a self-consistent mean-field approach, the
procedure is repeated until convergence to the desired accuracy is
achieved.}
\end{figure*}

In what follows, we introduce the new disorder self-energy framework,
outline its favorable scaling properties and present its application
to three distinct problems: (i) quantum criticality induced by chiral
disorder in the BDI symmetry class; (ii) scattering resonances in
a spin--orbit coupled ferromagnet; and (iii) disorder-enhanced $p$-wave
superconductivity in graphene.

\section{\textit{\emph{Spectral approach}}}

\subsection{Chebyshev expansion of the spectral function\label{subsec:A}}

To set the stage, let us briefly review the polynomial expansion of
the one-particle spectral function. Consider a general fermionic system
on a $d$-dimensional lattice described by a Hamiltonian with a bounded
spectrum, $\hat{H}$. To enable a decomposition of the spectral function
in terms of orthogonal polynomials, we first perform the following
linear transformation 
\begin{equation}
\hat{h}=(\hat{H}-E_{+}\mathbf{1})/E_{-}\,,\quad E_{\pm}=(E_{a}\pm E_{b})/2,\label{eq:canonical_transf}
\end{equation}
where $\mathbf{1}$ is the identity operator defined on the Hilbert
space of the lattice and $E_{a(b)}$ indicates the largest (smallest)
eigenvalue of $\hat{H}$. Note that this procedure maps the eigenvalues
of the Hamiltonian onto the canonical interval $\mathcal{I}=[-1,1]$.
Here, we employ Chebyshev polynomials of the first kind which are
particularly well-suited to approximate nonperiodic functions over
a finite interval on the real axis \citep{boydChebyshevFourierSpectral1989}.
The spectral operator $\hat{\mathcal{A}}(\varepsilon)=\delta(\varepsilon-\hat{h})$
associated with the rescaled Hamiltonian in Eq. (\ref{eq:canonical_transf})
can be decomposed into a Chebyshev series according to 
\begin{equation}
\hat{\mathcal{A}}(\varepsilon)=\,\sum_{n=0}^{\infty}\mathcal{\omega}_{n}(\varepsilon)\,T_{n}(\varepsilon)\,T_{n}(\hat{h})\,,\label{eq:spectral_operator}
\end{equation}
where $\mathcal{\omega}_{n}(\varepsilon)=2/[\pi(1+\delta_{n,0})(1-\varepsilon^{2})^{1/2}]$
is the weight function entering in the orthogonality relations and
$\varepsilon\equiv(E-E_{+})/E_{-}$ is the rescaled energy variable,
with $E$ the energy spectrum of the original Hamiltonian \citep{weisseKernelPolynomialMethod2006}.

The favorable convergence properties of the general-purpose Chebyshev
expansion in Eq. (\ref{eq:spectral_operator}) reflects its close
relation to the Fourier series made manifest by the identity $T_{n}(\varepsilon)=\cos(n\arccos\varepsilon)$.
We do not provide details on the underlying spectral theory but instead
refer the reader to Boyd's book \citep{boydChebyshevFourierSpectral1989}.
By virtue of the Chebyshev recurrence relations, 
\begin{align}
T_{0}(\hat{h})=\mathbf{1} & \,,\quad T_{1}(\hat{h})=\hat{h}\,,\label{eq:Cheb_rule_a}\\
T_{n+1}(\hat{h}) & =2\hat{h}T_{n}(\hat{h})-T_{n-1}(\hat{h})\,,\label{eq:Cheb_rule_b}
\end{align}
the matrix coefficients in Eq.~(\ref{eq:spectral_operator}) can
be obtained iteratively to any desired order. In a practical implementation,
it suffices to evaluate the scalar Chebyshev expansion coefficients
of the overlap $\mathcal{A}_{\phi\psi}(\varepsilon)=\langle\phi|\hat{\mathcal{A}}(\varepsilon)|\psi\rangle$.
The choice of basis functions $\phi,\psi$ defines the target function
of energy (see below). Because the recursive procedure {[}Eqs.\,(\ref{eq:Cheb_rule_a})-(\ref{eq:Cheb_rule_b}){]}
is highly stable, the reconstruction of the spectral function can
be carried out in principle with very high energy resolution $\gamma\approx1/M$
(in rescaled units), where $M$ is the truncation order. Such features
will be of key practical importance in the evaluation of the self-energy
operator, as we shall see briefly.

To illustrate the effectiveness of the spectral approach, we first
consider the local density of states (LDOS) at a site $i$, defined
as $\nu_{i}(\varepsilon)=\langle i|\hat{\mathcal{A}}(\varepsilon)|i\rangle$.
The LDOS is the simplest spectral target function of energy derived
from the spectral operator {[}note that it corresponds to a diagonal
element $\nu_{i}(\varepsilon)=\mathcal{A}_{ii}(\varepsilon)$ in the
lattice basis{]}, yet it contains rich information on electronic state
hybridization and scattering processes. From Eq.~(\ref{eq:spectral_operator}),
the $M$-order Chebyshev approximation to the LDOS is easily constructed
as
\begin{equation}
\nu_{i}^{M}(\varepsilon)=\sum_{n=0}^{M-1}\mu_{in}\,\mathcal{\omega}_{n}(\varepsilon)\,T_{n}(\varepsilon)\,,\label{eq:LDOS_approximant}
\end{equation}
with $\mu_{in}=\langle i|T_{n}(\hat{h})|i\rangle$ \footnote{The truncation of a polynomial expansion introduces spurious oscillations,
known as Runge phenomenon, arising from discontinuities or singularities
in the spectral function. These can be effectively damped out by convolution
with a suitable kernel function, $\hat{\mathcal{A}}_{\textrm{KPM}}(\varepsilon):=\int_{-1}^{1}d\varepsilon^{\prime}\hat{\mathcal{A}}(\varepsilon^{\prime})K(\epsilon,\epsilon^{\prime})\,$,
a strategy known as the kernel polynomial method \citep{weisseKernelPolynomialMethod2006}. }.

To retrieve the Chebyshev coefficients $\{\mu_{in}\}$, the Eqs. (\ref{eq:Cheb_rule_a})-(\ref{eq:Cheb_rule_b})
are iterated on the fly by exploiting the sparseness of minimal-basis
(real-space) representations. The problem is thus solved efficiently
by repeating the following steps. (i) Starting with the initial vectors
$|\psi_{0}^{i}\rangle:=|i\rangle$ and $|\psi_{1}^{i}\rangle:=\hat{h}|\psi_{0}^{i}\rangle$,
act iteratively with the rescaled Hamiltonian using the Chebyshev
rule $|\psi_{n+1}^{i}\rangle=2\hat{h}|\psi_{n}^{i}\rangle-|\psi_{n-1}^{i}\rangle$;
and (ii) At each step compute the overlaps $\mu_{ni}=\langle\psi_{0}^{i}|\psi_{n}^{i}\rangle$.
From the knowledge of the Chebyshev moments $\{\mu_{ni}\}$, the LDOS
in any desired energy range can be easily retrieved using Eq. (\ref{eq:LDOS_approximant}).
The overall computational cost is determined by the energy resolution
desired for the spectral reconstruction of the LDOS. For typical sparse
Hamiltonian matrices, the number of operations scales linearly with
respect to both the number of sites and number of moments $M$, which
makes the method particularly advantageous for single-electron problems
\citep{joaoKITEHighperformanceAccurate2020a}. The spectral approach
has been successfully employed to reveal the electronic structure
of a wide range of systems, including Anderson disorder models \citep{Schubert_05},
graphene with atomic defects \citep{Ferreira_11,Garcia_14,Cysne_16}
and disordered superconductor-normal metal interfaces \citep{Covaci_10}.
Other recent applications include the calculation of time-dependent
equilibrium Green's functions of superconductors \citep{PhysRevB.104.125405}
and dynamical structure factors in quantum spin chains \citep{Lado19}.
(For a review of early work, see Ref. \citep{weisseKernelPolynomialMethod2006}.)

\subsection{High-resolution self-energy calculation\label{subsec:B}}

Having laid out the basic principles underlying the efficient reconstruction
of the spectral function using Chebyshev polynomials, we now move
on to tackle the nontrivial problem of determining the self-energy
operator. The central objects of interest in this discussion are the
retarded lattice Green's functions and disorder self-energy 
\begin{align}
\hat{G}^{\eta}(\omega) & =\frac{1}{\omega-\hat{H}+i\eta},\;\;\hat{G}_{0}^{\eta}(\omega)=\frac{1}{\omega-\hat{H}_{0}+i\eta}\,,\label{eq:GF_total}\\
\hat{\Sigma}_{\textrm{dis}}^{\eta}(\omega) & =[\hat{G}_{0}^{\eta}(\omega)]^{-1}-\left\langle \hat{G}^{\eta}(\omega)\right\rangle {}^{-1}\,,\label{eq:self_energy}
\end{align}
where $\hat{H}_{0}$ is the clean Hamiltonian and $\eta$ plays the
role of an energy resolution (see below). As customary, the real-space
disorder is added to the Hamiltonian, $\hat{H}=\hat{H}_{0}+\hat{V}_{\textrm{dis}}$,
via random modifications of hopping amplitudes and on-site energies.
Of particular interest is the wavevector ($\mathbf{k}$) dependence
and internal (orbital and spin) structure of $\hat{\Sigma}^{\eta}(\omega)$.
It is natural to ask whether the accurate LDOS polynomial scheme can
be extended to the matrix inverse problem posed by Eq.~(\ref{eq:self_energy}).
An immediate stumbling block is simply that the inversion procedure
is prone to loss of accuracy, particularly for weak disorder, due
to the parametrically small difference between the clean and disordered
Green's functions. As shown below, such a hurdle can be overcome by
devising a spectral algorithm that accurately reconstructs the Green's
function projected onto the local basis elements, thereby effectively
mapping the problem into an $N\times N$ matrix inversion, where $N$
is the number of bands. A more subtle issue concerns the stability
of the self-energy with respect to the broadening scheme of the Green's
function of finite systems \citep{Thouless_1981,Imry2002}. For many
problems of interest, one must be able resolve the fine structure
of the disordered Green's function and thus $\eta$ must be comparable
or smaller than the self-energy itself. In practical terms, $\eta$
is bounded from below by the mean level spacing and a careful convergence
analysis is required to obtain sensible thermodynamic results.

We posit that the Chebyshev polynomial-based spectral approach is
well suited to overcome the above obstacles, since it allows for real-space
calculations with high accuracy and well-defined energy resolution.
The standard strategy to mimic the broadening in Eq.~(\ref{eq:GF_total})
is to formally expand the Green's function $\hat{G}\left(\omega\right)=\lim_{\eta\rightarrow0^{+}}\hat{G}^{\eta}\left(\omega\right)$
in Chebyshev polynomials and regularize the resulting spectral series
through convolution with a Lorentzian kernel \citep{Wolf_14,Holzner_11}.
Here we propose an alternative approach based on the direct expansion
of the broadened Green's function $\hat{G}^{\eta}(\omega)$ into Chebyshev
polynomials of the first kind.

Let us start by defining the rescaled Green's function operator 
\begin{equation}
\hat{\mathcal{G}}(z)\equiv E_{-}\hat{G}^{\eta}(\omega)=(z-\hat{h})^{-1}\,,\quad z=\varepsilon+i\gamma\label{eq:rescaled_GF}
\end{equation}
where we have used Eq.~(\ref{eq:canonical_transf}) and defined $\gamma=\eta/E_{-}$.
This rescaled Green's function then admits the following exact decomposition

\begin{equation}
\hat{\mathcal{G}}\left(z\right)=\sum_{n=0}^{\infty}c_{n}(z)\,T_{n}(\hat{h})\,,\quad c_{n}(z)=a_{n}\frac{[z-f(z)]^{n}}{f(z)}\,,\label{eq:CPGF}
\end{equation}
where $a_{n}=2/(1+\delta_{n,0})$ and $f(z)=i\sqrt{1-z^{2}}$ \citep{Braun_14,ferreiraCriticalDelocalizationChiral2015}.
The main advantage of this variant of the familiar kernel polynomial
method is that the energy levels are probed with known uniform resolution
over the entire spectral range $\varepsilon\in[-1,1]$. The energy
resolution can be made as high as desired by a judicious truncation
of Eq.~(\ref{eq:CPGF}); as a rule of thumb, the required number
of Chebyshev iterations is $M_{\gamma}\approx[1/\gamma]$. This spectral
scheme, combined with an efficient real-space implementation, will
allow us to resolve the fine structure of the quasiparticle self-energy
in large systems containing multi billions of orbitals, a task that
has remained elusive thus far. Such a capability is key to exploring
topological transitions and disordered systems at quantum criticality,
where spectral convergence is already challenging at the level of
considerably simpler average density of states \citep{Hafner_14,ferreiraCriticalDelocalizationChiral2015,Pires_21}.

With regards to the disorder averaging procedure {[}Eq. (\ref{eq:self_energy}){]},
a brief discussion is in order. For the cases of interest here, the
$\mathbf{k}$-space Green's function is found to exhibit self-averaging
behavior. Thus, the disorder self-energy of a single large sample
is representative of the whole ensemble, greatly reducing the computational
cost related to configurational averaging. In practical terms, the
self-averaging property of the disorder self-energy is demonstrated
numerically by analyzing the scaling of self-energy fluctuations with
the system size. For the interested reader, we provide analytical
proofs for two common classes of problems in Appendix \ref{Appendix_A_self_averaging}.
Our findings suggest that self-averaging is a universal property of
the $\mathbf{k}$-space disorder self-energy, which will be explored
in future work.

To facilitate the evaluation of the disorder self-energy, let us introduce
the orthogonal basis set of plane-wave states of wavevector \textbf{k,}
$\{|\mathbf{k},\alpha\rangle\}$, where $\alpha=1...N$ labels the
internal quantum numbers of the electronic system. We first evaluate
the Green's function matrix elements, $\mathcal{G}_{\alpha\beta}\left(\mathbf{k},z\right)=\langle\mathbf{k},\alpha|\mathcal{\hat{G}}\left(z\right)|\mathbf{k},\beta\rangle$,
with the desired spectral resolution. The projected Green's function
Chebyshev moments, $\mu_{\alpha\beta}^{n}(\mathbf{k})=\langle\mathbf{k},\alpha|T_{n}(\hat{h})|\mathbf{k},\beta\rangle$,
are then computed with the recursive scheme described in Sec. \ref{subsec:A}.
Specifically, we compute the $N\times N$ overlap matrix
\begin{figure*}
\begin{centering}
\includegraphics[width=1\textwidth]{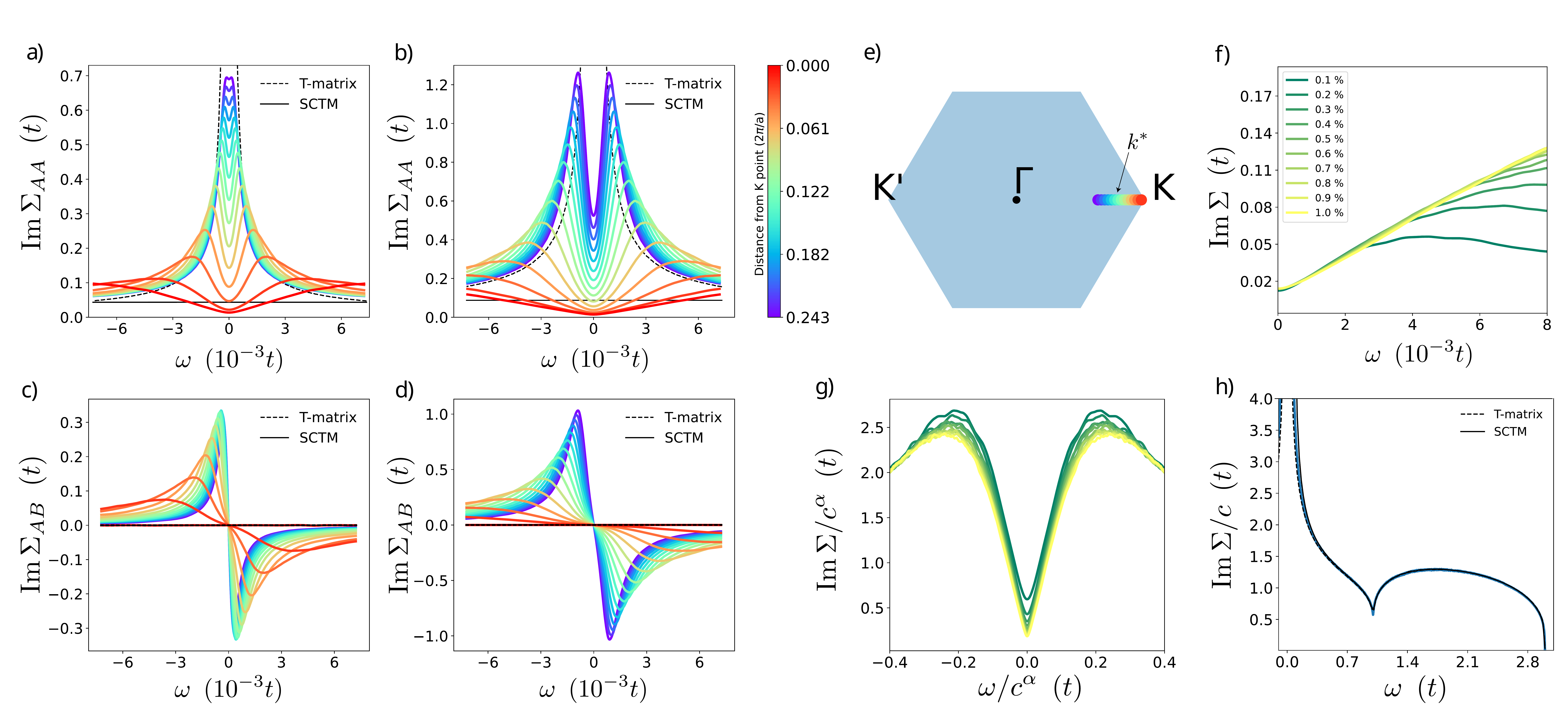}
\par\end{centering}
\caption{\label{fig:02}Momentum-resolved disorder self-energy operator of
graphene. a) Imaginary part of the diagonal component $\Sigma_{AA}(\mathbf{k},\omega)$
as a function of energy for vacancy concentration of $0.3\%$ calculated
for several $\mathbf{k}$ points with a resolution $\eta=3\times10^{-4}t$
($\simeq0.8$ meV). The dashed (solid) black curve represents the
$T$-matrix (SCTM) approximation. b) same as a) but for a concentration
of $1\%$. (c-d) Same as (a-b) but for the off-diagonal component
of the self-energy matrix $\Sigma_{AB}(\mathbf{k},\omega)$. Both
the T-matrix and the SCTM are zero in these graphs. e) $\mathbf{k}$-path
corresponding to the color scale in a), b), c) and d). f) Diagonal
component of the self-energy matrix at the Dirac point ($\mathbf{k}=0$)
at selected concentrations and small energies. g) Collapsing of data
using the proposed ansatz $\textrm{Im}\,\Sigma(\omega)=c^{\alpha}f(\omega c^{\alpha})$
with $\alpha=0.56\pm0.02$. h) same as f) but normalized to $c$ and
for the spectral region from zero to the band edge.}
\label{vacancies graphene}
\end{figure*}
\begin{equation}
\mu_{\alpha\beta}^{n}(\mathbf{k})=\langle\mathbf{k},\alpha|\psi_{n}^{\beta\mathbf{k}}\rangle\;\quad(n=0,...,M_{\gamma}-1)\,\label{eq:overlap_matrix}
\end{equation}
along the desired $\mathbf{k}$ paths, with $|\psi_{\beta\mathbf{k}}^{n}\rangle$
the $n$-th Chebyshev vector obtained by successive applications of
Eqs.~(\ref{eq:Cheb_rule_a})-(\ref{eq:Cheb_rule_b}) with the initial
vector $|\psi_{\beta\mathbf{k}}^{0}\rangle=|\mathbf{k},\beta\rangle$.
This will afford us with several computational advantages, in particular,
the possibility to tackle very large systems (note that only 3 vectors
of the Hilbert space dimension need to be stored for each $\mathbf{k}$
point). Next, the $N\times N$ $\mathbf{k}$-space Green's function
is reconstructed using the truncated expansion

\begin{align}
\mathcal{G}_{\alpha\beta}\left(\mathbf{k},z\right) & \cong\sum_{n=0}^{M_{\gamma}-1}c_{n}(z)\mu_{\alpha\beta}^{n}(\mathbf{k})\,.\label{eq:GF_self_en_dis_av}
\end{align}
With a suitable choice of $M_{\gamma}$, Eq.~(\ref{eq:GF_self_en_dis_av})
yields numerically exact results for the broadened lattice Green's
function within machine precision. Furthermore, the dependence of
the Green's function with the resolution parameter (up to $\gamma\approx1/M_{\gamma}$)
can be retrieved without having to recalculate the moments $\mu_{\alpha\beta}^{n}(\mathbf{k})$,
which allows for assessing convergence on the fly. The momentum-space
Green's function of the clean system can be calculated along the same
lines (or simply through direct diagonalization of the corresponding
Bloch Hamiltonian). The final step is a simple inversion of two $N\times N$
matrices,

\begin{equation}
\Sigma_{\textrm{dis}}^{\eta}\left(\mathbf{k},\omega\right)=[G_{0}^{\eta}\left(\mathbf{k},\omega\right)]^{-1}-[G^{\eta}\left(\mathbf{k},\omega\right)]^{-1},\label{eq:se_formula}
\end{equation}
which reconstructs the momentum-space self-energy in a fully non-perturbative
manner. This simple inversion of the lattice Green's function $G^{\eta}$
is justified because, for self-averaging systems, it is diagonal in
$\mathbf{k}$-space (see Appendix \ref{Appendix_A_self_averaging}
for a detailed discussion). The procedure may, of course, be repeated
for any number of samples in order to recover the disorder-averaged
Green's function in the traditional sense if self-averaging is not
obvious {[}note that, in general, $\Sigma^{\eta}$ and $G^{\eta}$
in Eq.~(\ref{eq:se_formula}) should be understood as the disorder-averaged
operators $\Sigma^{\eta}=\langle\Sigma^{\eta}\rangle$ and $G^{\eta}=\langle G^{\eta}\rangle$.{]}.
Our self-energy spectral framework is summarized in Fig. \ref{fig:01}.

Before closing this subsection, we briefly comment on the computational
cost associated with the calculation of the self-energy. Because the
recursive algorithm exhibits polynomial complexity {[}see Eq.~(\ref{eq:overlap_matrix})
and text therein{]}, one can extract the self-energy operator of considerably
large systems with a small computational cost. For example, a single-orbital
tight-binding model on a square-lattice with real first-neighbor hoping,
$10^{3}\times10^{3}$ sites and uncorrelated on-site disorder, requires
1 GB RAM and 1 core hours to reconstruct the self-energy operator
at a fixed $\mathbf{k}$ point and disorder realization with $2000$
Chebyshev iterations ($\eta\approx0.001$ in units of bandwidth).
Significantly more complex problems can be tackled by optimizing the
parallelization efficiency of matrix-vector multiplications using
an adaptive real-space domain decomposition algorithm as implemented
in the \texttt{KITE} package \citep{joaoKITEHighperformanceAccurate2020a}
(see Appendix \ref{Appendix_B_domain_decomp} for more details). Such
a strategy was adopted in recent works reporting accurate studies
of the effect of short-range disorder on the nodal density of states
of topological semimetals \citep{Goncalves_20,Pires_21}.

\section{Results}

\subsection{\textit{\emph{Gade singularity of graphene: $\mathbf{k}$-dependence}}\textit{\label{subsec:Example_1_Gade} }}

We now turn to the presentation of the results obtained with the spectral
method introduced above. As a first case study, we consider graphene
with vacancy defects \citep{RMP_graphene,Ugeda10,Nanda_2012}. Site
dilution in a honeycomb layer provides an intriguing example of a
random-hopping system which exhibits anomalous quantum critical behavior
stemming from sublattice (chiral) symmetry \citep{gadeReplicaLimitModels1991,gadeAndersonLocalizationSublattice1993}.
Chiral-symmetric disorder induces critically delocalized states at
the band center {[}so-called zero energy modes (ZEMs){]} characterized
by a Gade singularity in the density of states \citep{ostrovskyDensityStatesTwoDimensional2014,Motrunich_02,Mudry_03,Hafner_14,Sbierski_20}.
Moreover, quantum transport simulations indicate that dilute ZEMs
can overcome Anderson localization in an infinite system, with conductivity
pinned to the $T=0$ ballistic value $\sigma_{0}=(4/\pi)e^{2}/h$
irrespective of the vacancy concentration \citep{ferreiraCriticalDelocalizationChiral2015}.
The puzzling behavior of ZEMs has been attributed to unusually strong
nonperturbative quantum interference effects (beyond standard field-theoretic
treatments), but direct evidence in $\mathbf{k}$-space has remained
elusive. In order to model the electronic structure of graphene with
randomly distributed vacancies, we resort to a minimal tight-binding
model for spinless fermions on the honeycomb lattice
\begin{equation}
\hat{H}=-\sum_{\langle i,j\rangle}t_{ij}c_{i}^{\dagger}c_{j}\,,\label{eq:graphene}
\end{equation}
where $c_{i}^{\dagger}$ ($c_{i}$) adds (removes) an electron to
the $i$-th site and $\langle i,j\rangle$ denotes nearest-neighbor
pairs of sites. Furhermore, $t_{ij}=t$ if $i$ and $j$ are both
undiluted sites, otherwise $t_{ij}=0$. The model in Eq.\,(\ref{eq:graphene})
possesses both time-reversal ($\mathcal{T}^{2}=+1$) and particle-hole
($\mathcal{C}^{2}=+1$) symmetries, placing it in the chiral orthogonal
(BDI) class of the topological classification \citep{Altland_97}.
Thus, random vacancies can be viewed as topological point defects
which preserve the underlying non-spatial symmetries of the host crystal,
most notably its chiral-sublattice symmetry, i.e. $\mathcal{S}\hat{H}\mathcal{S}=-\hat{H}$.
Here, $\mathcal{S\equiv\mathcal{T}\mathcal{C}}=\sigma_{z}$, with
$\sigma_{z}$ the diagonal Pauli matrix defined in the $A$-$B$ sublattice
space.

Previous work has calculated the vacancy-induced self-energy in the
continuum limit of Eq.\,(\ref{eq:graphene}) by means of the self-consistent
$T$-matrix (SCTM) approximation \citep{Ostrovsky_06}. The SCTM can
be evaluated analytically both near the band center and far from it,
yielding a scalar self-energy of the form $\text{Im}\:\Sigma(\omega)\propto\Gamma$
for $|\omega|\ll\Gamma$ ($\text{Im}\:\Sigma(\omega)\propto-c/|\omega|\log\left(|\omega|\right)$
for $|\omega|\gg\Gamma$), where $\Gamma=\Lambda\sqrt{-c/\log\left(c\right)}$
with $c$ the vacancy concentration and $\Lambda$ a suitable ultraviolet
cutoff \citep{Ostrovsky_06}. Although the SCTM provides a faithful
description of the problem in the semiclassical regime ($\omega\gg\Gamma$),
it cannot reproduce neither the Gade singularity in the density of
states \citep{Hafner_14}, nor the anomalous behavior in the conductivity
of ZEMs \citep{ferreiraCriticalDelocalizationChiral2015}, because
it neglects quantum coherent multiple scattering. Moreover, the continuum
model treats vacancies as $\delta$-peak potentials centered at random
positions, which results in a structureless (i.e. $\mathbf{k}$-independent)
self-energy operator at all energies. To overcome these limitations,
we use our high-resolution spectral approach to map out the momentum-space
self-energy of the lattice model. For definiteness, we focus on the
case of compensated vacancies equally distributed on both sublattices.
The Chebyshev-polynomial-based reconstruction of the self-energy is
carried out on very large systems with up to $10^{9}$ sites and $M=2^{16}=65536$
moments, giving us unprecedented access to sub-meV resolution over
the entire ($\omega$,$\mathbf{k}$)-parameter space. The disorder
self-energy is projected onto the sublattice space according to $\Sigma_{\alpha\beta}^{\eta}(\mathbf{k},\omega)=\langle\mathbf{k},\alpha|\hat{\Sigma}^{\eta}(\omega)|\mathbf{k},\beta\rangle$,
with $\alpha,\beta=A,B$. Homogeneity implies $\Sigma_{AA}=\Sigma_{BB}$
and $\Sigma_{AB}=\Sigma_{BA}$. Thus it suffices to find the $AA$
and $AB$ elements. Furthermore, particle-hole symmetry implies $\Sigma_{AA}(\omega)=-\Sigma_{AA}^{*}(-\omega)$
and $\Sigma_{AB}(\omega)=\Sigma_{AB}^{*}(-\omega)$. The self-averaging
property combined with the very large lattice sizes provides the additional
advantage that a single disorder landscape is required for our purposes
(see Appendices \ref{Appendix_A_self_averaging} and \ref{Appendix_C_spectral_conv}
for additional details on the self-averaging property and spectral
convergence, respectively).

We focus the subsequent discussions on the imaginary part of the self-energy
operator, which encodes the quasiparticle lifetime. As a reference
point, we calculate the $T$-matrix and SCTM self-energies using the
lattice Green's function of the clean model. The fully converged results,
summarized in Fig.~\ref{fig:02}, contain a number of surprising
findings. First, the disorder self-energy shows a strong $\mathbf{k}$-dependence
in the vicinity of the band center, where the Gade singularity is
located. Moreover, the point defects endow the self-energy with a
nonzero off-diagonal component in that same region. Second, the concentration
dependence of the self-energy exhibits anomalous scaling at the lowest
energies, where a rich crossover between the quantum critical regime
near $\omega=0$ and the pure semiclassical regime at high energies
can be seen. All these features are missing from the SCTM approximation
and, as argued below, provide fingerprints of the conjectured strong
nonperturbative quantum interference effects induced by ZEMs in graphene.

A close up of the self-energy matrix elements around the Gade singularity
are shown in Figs.~\ref{fig:02}(a)-(d). The observed fine structure
is confined to a narrow window of width $\delta\approx2tc^{0.6}$.
The twin peaks in the diagonal elements $\Sigma_{AA(BB)}$ borne out
by our high-resolution data become visibly sharper as one moves away
from the $K$ point along the path indicated in Fig. \ref{fig:02}(e),
with other paths showing similar behavior. Let us note that the distance
between these peaks decreases, while their height increases, as we
move away from the K point, such that the curves approach that of
the $T$-matrix approximation. However, whereas the latter is structureless
and diverges as $\textrm{Im}\,\Sigma_{AA}\sim-1/\left|\omega\right|\log\left|\omega\right|$,
the numerically exact self-energy is strongly $\mathbf{k}$-dependent
and bounded in the vicinity of the Dirac ($K$) point. The SCTM approximation
effectively smears the bare $T$-matrix result (thus removing the
divergence at $\omega=0$), however, neither approach captures the
nonperturbative behavior seen at low energy. Strikingly, the $\omega=0$
self-energy at the Dirac point approaches zero as $\eta\rightarrow0$
at all concentrations, which implies a divergent elastic scattering
time at the $K$ point (see Appendix \ref{Appendix_C_spectral_conv}
for a scaling analysis). We speculate that the exceedingly large quasiparticle
lifetime protects ZEMs against backscattering, thus providing a new
insight into the ``mysterious'' ZEM resilience observed in large-scale
simulations of the dc conductivity \citep{ferreiraCriticalDelocalizationChiral2015}.
We stress that the favorable scaling of our method is crucial to uncover
the fine features of the self-energy, a task which requires both very
large samples and fine resolution $\eta\apprle\delta$.

We now briefly discuss the sublattice-coherence effects encoded in
the off-diagonal elements of the $\mathbf{k}$-space self-energy,
$\Sigma_{AB(BA)}$. Figure~\ref{fig:02}(d) shows that the off-diagonal
component acquires the form of a symmetric Fano resonance with sizeable
amplitude away from the Dirac point {[}e.g., $\textrm{Im}\Sigma_{AB}\approx0.3t$
for $c=0.003$ and $\mathbf{k}=\mathbf{k}^{*}\simeq0.76|\mathbf{K}|\hat{x}$;
see Fig. \ref{fig:02}(e){]}. Within the $T$-matrix or SCTM approach,
the off-diagonal self-energy is identically zero, which shows that
the $AB$-sublattice coherence captured by our real-space spectral
method is inherently a nonperturbative effect resulting from coherent
multi-impurity scattering events.

\begin{figure*}
\begin{centering}
\includegraphics[width=1\textwidth]{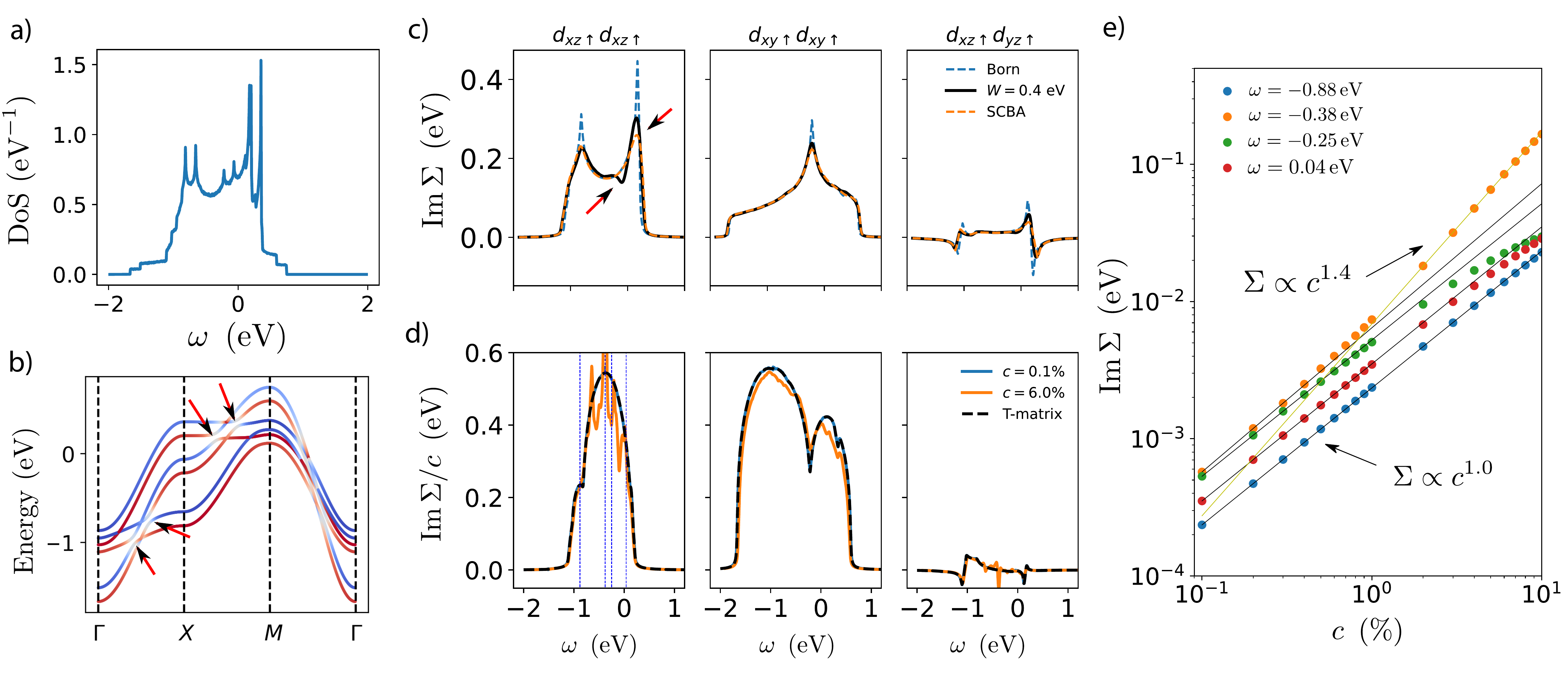}
\par\end{centering}
\caption{a) Average density of states of the interfacial layers of a SRO superlattice.
b) Band structure (colors represent the spin expectation value $\left\langle s_{z}\right\rangle $
with blue/red for negative/positive values). The red arrows indicate
the presence of nodal loops in the spectrum. c) Imaginary part of
three components of the self-energy matrix. The red arrows point to
the regions of largest discrepancy between SCBA and the exact self-energy.
d) Vacancy-induced self-energy normalized to the concentration and
comparison with the $T$-matrix result for $c=0.1\%$ and $c=6\%$.
e) Imaginary $d_{xz}\uparrow,d_{xz}\uparrow$-components of the self-energy
at fixed energy (blue dashed lines in (d)) for several concentrations
of vacancies. These simulations were carried out using computational
domains with a total of $2.5\times10^{7}$ sites.}
\label{fig03}
\end{figure*}

In order to unveil the extent of quantum coherent scattering effects,
we extracted the vacancy concentration dependence of the self-energy
over the entire spectrum. First, we note that the self-energy converges
to its bare $T$-matrix form in the high electronic density regime,
where chiral symmetry is absent and the system falls under the standard
orthogonal Wigner-Dyson class; see Fig. \ref{fig:02}(h). Now, there
are two relevant (concentration dependent) energy scales in this problem:
(i) the energy at which the results start to diverege significantly
from the T-matrix result, $\Gamma_{T}\propto c^{0.2}$, and; (ii)
the energy $\Gamma_{k}<\Gamma_{T}$ at which the $\mathbf{k}$-dependency
becomes very strong, which scales roughly as $c^{0.6}$. For energies
$|\omega|\gg\Gamma_{k}$, the self-energy is $\mathbf{k}$-independent
and, for larger energies still, when $\left|\omega\right|\gg\Gamma_{T}$,
it matches the simple $T$-matrix expression $\Sigma(\omega)\approx-ic/\left|\omega\right|\log\left|\omega\right|$.
The linear dependence upon the concentration is the signature of the
semiclassical regime, where single-impurity scattering events dominate.
However, much more interesting is the vicinity of the BDI quantum
critical point, where the perturbative picture breaks down \citep{Motrunich_02,Mudry_03,ostrovskyDensityStatesTwoDimensional2014,Hafner_14,Sbierski_20,ferreiraCriticalDelocalizationChiral2015}.
Near the Dirac nodes ($\omega\rightarrow0$), the self-energy becomes
independent of the concentration. A close up of the concentration
dependence at the Dirac point is shown in Fig. \ref{fig:02}(f). This
anomalous behavior is found to occur for energies that are within
the energy window where the $\mathbf{k}$ dependence is the strongest,
which indicates that both effects have origin in high order (multiple
impurity) scattering processes. We note that the onset energy of this
anomalous behaviour increases with the increase of vacancy concentration.
This means that the manifestations of quantum criticality highlighted
here can be pushed towards experimentally accessible energy scales
in sufficiently disordered samples. Of equal interest is the intermediate
energy regime {[}Fig. \ref{fig:02}(g){]}, where considerable deviations
from the semiclassical picture are also apparent. Here, the $K$-point
self-energy is found to follow the scaling law $\textrm{Im}\,\Sigma_{AA}\left(0,\omega\right)=c^{\alpha}f\left(\omega c^{\alpha}\right)$,
see Fig. (\ref{vacancies graphene}d), where $f(x)$ is independent
on $c$ and $\alpha=0.56\pm0.02$.

\subsection{Spin--orbit effects in ferromagnets:\textit{ SrRu$O_{3}$\label{subsec:Example_2_SRO}}}

Next, we illustrate the versatility of our approach by computing the
disorder self-energy matrix of a spin--orbit-coupled ferromagnetic
metal. The model system chosen for this study is the itinerant ferromagnet
SrRuO$_{3}$ (SRO) \citep{SRO_RevModPhys.84.253}, a well-studied
oxide material that features, among other things, momentum-space monopoles
of Berry curvature and interface-driven chiral spin textures \citep{SRO_Science_03,SRO_topology_1,SRO_topology_2,SRO_topology_3,SRO_topology_4}.
Our focus here is on the spin-polarized two-dimensional electron gases
that are formed in SrRuO$_{3}$ embedded in a SrTiO$_{3}$ matrix
\citep{PhysRevX.9.011027}. Complex-oxide superlattices have attracted
widespread interest because they provide a platform to engineer metallic
states with unusual ferroelectric properties \citep{Jang886,Mao2021,Meng2019,Bangjae2020}.
However, the effect of disorder upon their complex interfacial behavior
remains largely unexplored.

To model an emergent spin-polarized electron gas in a SrRuO$_{3}$/SrTiO$_{3}$
superlattice (which is known to be confined to the $4d$ orbitals
of the Ru atoms in the SRO layers \citep{PhysRevLett.108.107003}),
we employ a first-principles parameterized multi-orbital TB model
as developed in Ref. \citep{SRO_topology_3}. The details of the TB
model, which accurately describes the predominant Ru $t_{2g}$-bands
near the Fermi level and also includes spin--orbit coupling (SOC),
are provided in Appendix \ref{Appendix_D_SRO-1.Hamiltonian}. Our
primary aim is to understand whether the self-energy operator acquires
a nontrivial matrix structure. To this end, we investigate the quasiparticle
self-energy generated by on-site disorder and point-like defects (vacancies).
In the former, the random on-site energies within each unit cell in
real space $\{\varepsilon_{i}\}$ are taken from a box distribution
\begin{equation}
p(\varepsilon_{i})=\begin{cases}
1/\mathcal{W}, & |\varepsilon_{i}|\le\mathcal{W}/2\\
0, & \textrm{otherwise}
\end{cases}\,,\label{eq: SRO - distribution Anderson}
\end{equation}
where $\mathcal{W}$ defines the disorder strength. With this choice,
the on-site disorder is locally correlated since all the $4d$-orbitals
inside a unit cell experience the same potential. Different choices
are possible, but this simple prescription will be sufficient to illustrate
the nontrivial role of disorder in this class of oxide materials.

We first briefly describe the electronic structure of the spin-polarized
gas formed at the interfacial layers of SRO superlattices. Figure
\ref{fig03}(a) shows the average density of states and Fig. \ref{fig03}(b)
shows the band structure along the path $\Gamma XM\Gamma$. In the
absence of SOC, nodal loops are formed when the minority and majority
bands intersect (see arrows). The majority and minority spin bands
are hybridized when SOC is included, leading to a modulation of the
equilibrium $\mathbf{k}$-space spin-polarization density and an enhanced
Berry curvature near the avoided anticrossings \citep{SRO_Science_03}. 

We now discuss the quasiparticle self-energy at the $\Gamma$ point,
$\Sigma^{\eta}(\omega)\equiv\Sigma^{\eta}(\mathbf{k=\Gamma},\omega)$
--- other $\mathbf{k}$ points behave similarly and hence are not
further discussed here. The energy resolution in the Chebyshev polynomial
expansion {[}Eq. (\ref{eq:CPGF}){]} is set to $\eta=1$ meV. As it
turns out, excellent spectral convergence is achieved after $N\simeq16384$
iterations. Our results summarized in Figs. \ref{fig03}(c)-(d) disclose
a rich self-energy structure. For conciseness, only the dominant matrix
elements are shown (there are $18$ nonzero matrix elements in total;
see Appendix \ref{Appendix_D_SRO - 2.Other-matrix-elements}). These
results counter conventional wisdom, which posits that the self-energy
has essentially a scalar structure $\Sigma^{\eta\rightarrow0^{+}}=-i\,\Gamma$.
The scalar component of the self-energy is dominant in relatively
simple systems, such as graphene with point defects as discussed earlier
(Sec. \ref{subsec:A}), which, appropriately far from the Gade singularity,
displays an essentially scalar structure. However, the current example
clearly illustrates that all symmetry-allowed matrix elements of the
self-energy are generally nonzero, provided the existence of one-body
interactions coupling different degrees of freedom in $\hat{H}_{0}$.
In fact, the disorder self-energy shares its matrix structure with
the clean Hamiltonian. In particular, the SOC term in $\hat{H}_{0}$
is responsible for the nonzero spin-flip components of the self-energy
discussed below. The types of impurity potential and disorder statistics
also play a crucial role in determining the self-energy matrix structure.
Because we have considered a uniform on-site potential across all
4d-orbitals on a given impurity site {[}Eq. (\ref{eq: SRO - distribution Anderson}){]},
the impurity scattering effectively acts as a source of correlation
between every orbital. The disorder correlation is thus responsible
for the emergence of off-diagonal matrix elements at the lowest Born
order in the diagrammatic expansion (such as $\Sigma_{yz\uparrow,xz\uparrow}$),
which would otherwise be forbidden. For further details, we direct
the reader to Appendix \ref{Appendix_E_disorder_correlations}.

It is interesting to contrast the results for random on-site disorder
against a standard diagrammatic calculation performed at the bare
Born approximation (BA) and self-consistent Born approximation (SCBA)
levels. We find that for weak disorder ($W\lesssim50$ meV), the numerically
exact, BA and SCBA self-energy are all in excellent agreement. For
intermediate disorder strengths $W\approx0.5$ eV, the simple BA is
no longer able to provide a satisfactory approximation to the self-energy.
While the SCBA produces a better agreement, it is still unable to
capture some of the finer details of the self-energy that our method
captures (red arrows in Fig. \ref{fig03}c). This is to be expected,
since, for strong disorder, quantum interference corrections are ubiquitous
and cannot be captured by the SCBA. Interestingly, the combination
of disorder correlation between the spins and a considerable (45 meV)
SOC term induces a substantial spin-flip matrix element $d_{xz,\uparrow}d_{xy,\downarrow}$
(see Appendix \ref{Appendix_D_SRO - 2.Other-matrix-elements} for
the other matrix elements).

For the second model of disorder, we use vacancies with concentration
$c$ and compare our results to the $T$-matrix approximation {[}Fig.
\ref{fig03}(d){]}. At low concentrations ($c\sim0.1\%$), the $T$-matrix
approximation is in complete agreement with our results, but starts
to break down at higher concentrations ($c\sim5\%$). To see this
more clearly, we show the self-energy as a function of the concentration
for fixed energies in Fig. \ref{fig03}(e). For low concentrations,
the self-energy is proportional to the concentration of impurities
in accord with the $T$-matrix result, but at higher concentrations,
we start to see discrepancies which scale as $\sim c^{1.4}$ near
the peak, signaling the onset of nonperturbative disorder corrections.
We note that such peaks cannot be attributed to van Hove singularities
because they are absent at low defect concentration, and there is
no correlation between the position of the peaks and the position
of the singularities. We attribute them to resonances induced by multi-vacancy
clusters, which only start to form at higher defect concentrations.

The renormalization of quasiparticles with a non-scalar self-energy,
as predicted here, is expected to strongly impact the response of
materials to external perturbations. For example, recent theoretical
studies have alluded to robust spin--orbit scattering mechanisms
underlying the extrinsic generation of spin Hall currents and current-induced
spin polarization, which can be traced back to the matrix structure
of the disorder self-energy in spin--orbit coupled materials \citep{Offidani_17,offidaniAnomalousHallEffect2018,Sousa_2020}.
On a fundamental level, the self-energy is connected to the four-point
vertex functions of linear response theory through exact symmetry
relations known as Ward identities \citep{PhysRevB.64.115115,PhysRevB.77.041201,Milletari_17}
and thus the knowledge of all its matrix elements is essential to
obtain physically sensible transport equations. Because our approach
provides a systematic way to accurately evaluate the disorder self-energy
of arbitrarily complex model Hamiltonians, regardless of the type
and strength of disorder, it could provide new insights into the array
of rich interfacial magnetic phenomena beyond the reach of diagrammatic
calculations \citep{RevModPhys.89.025006}.

\subsection{\textit{\emph{Disorder-enhanced p-wave superconductivity}}\textit{\label{subsec:Example_3_pwave}}}

While our discussion so far has focused on lattice models with conventional
quasiparticles, it is straightforward to generalize our approach to
other condensed phases. As a final application, we employ our computational
machinery to map out the mean-field phase diagram of a dirty superconductor.
For definiteness, we focus on monolayer graphene, whose leading doping-dependent
superconducting instabilities include chiral $p$-wave pairing states
\citep{uchoaSuperconductingStatesPure2007a,Faye_15}. Chiral superconductivity
has caused great excitement because it provides a platform to realize
Majorana zero modes that are insensitive to local perturbations, and
thus can be used to construct topological qubits \citep{ivanovNonAbelianStatisticsHalfQuantum2001a}.
Typically, disorder is detrimental for unconventional (non $s$-wave)
superconductivity due to the breakdown of Anderson's theorem when
the impurities violate the pairing symmetry thus acting as pair breakers
\citep{Anderson_59,PhysRevB.48.4219,PhysRevB.56.7882,PhysRevB.71.094516}.
However, there are exceptions to this rule (e.g., in $d$-wave cuprates,
disorder is known to enhance the critical temperature through the
appearance of superconducting islands around the impurities \citep{lerouxDisorderRaisesCritical2019}).
In this respect, the unusual electronic properties of graphene open
up interesting possibilities. For example, it is known that the addition
of scalar impurities in charge-neutral graphene has the counterintuitive
effect of enabling conventional superconductivity for weak attractive
interactions, while leading to a suppression of superconductivity
in the strong attraction regime \citep{PhysRevB.87.174511,PhysRevB.90.094516}.
On the other hand, the phase diagram at finite doping, where superconductivity
is expected to develop more easily due to a nonzero single-particle
density of states, is far less explored. How is the doped graphene's
ability to form superconducting states affected by impurity scattering?
Can one tune the competition between different pairing states by tailoring
the impurity potential (e.g., using adatoms adsorbed on particular
lattice positions)? Here, we make a start on addressing these questions
by computing the superconducting gaps at finite charge carrier density
in the presence of disorder and competing order parameters. In the
following, we demonstrate that $p$-wave superconductivity can be
substantially enhanced by the presence of bond disorder (e.g. generated
by gauge fields due to strain or adatoms at bridge sites).

\begin{figure*}
\begin{centering}
\includegraphics[width=1\textwidth]{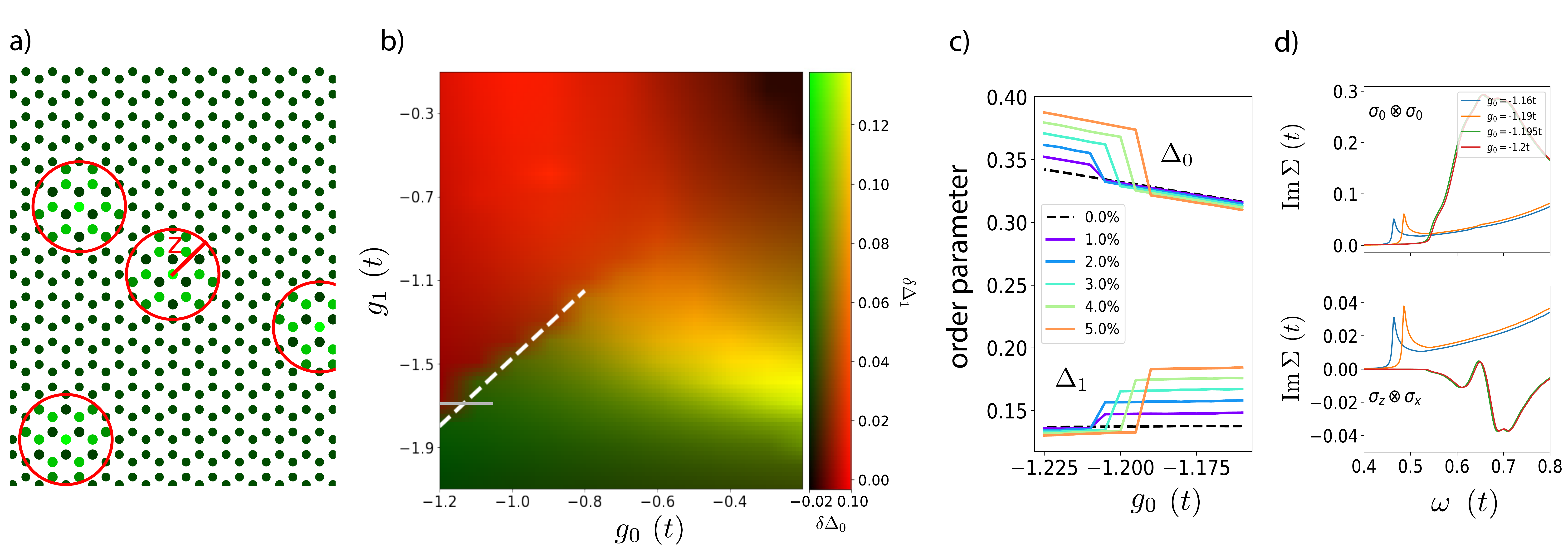}
\par\end{centering}
\caption{a)\textcolor{black}{{} Implementation: Density map of order parameter
$\Delta_{0}=f_{0}(\mathbf{r})$ around impurity positions in the honeycomb
lattice. Lighter shades of green indicate larger $|\Delta_{0}|$.
The spatial modulation of $\Delta_{0}$ is confined to a circle of
radius $z$ (shown in red) and forced to be identical in every circle.
}b) Variation in the s- and p-wave order parameters ($\Delta_{0}$
and $\Delta_{1}$) due to the presence of impurities at a concentration
$c=0.05$ and chemical potential $\mu=0.4t$. c) Order parameters
for fixed $g_{1}$ as a function of the concentration, across the
transition line. d) Components of the self-energy for four points
in the phase diagram in the $g_{1}=-1.7t$ line. For all our simulations,
we used systems with $10^{6}$ atoms and $2048$ Chebyshev polynomials.}
\label{fig04}
\end{figure*}

Chebyshev-Bogoliubov-de Gennes formalism---The algorithm we use is
a variant of the Chebyshev-Bogoliubov-de Gennes algorithm proposed
in Ref.~\citep{covaciEfficientNumericalApproach2010}, adapted here
to reconstruct real-space Green's functions by means of the exact
spectral decomposition {[}Eq. (\ref{eq:CPGF}){]} throughout the mean-field
self-consistency cycle; see Fig.\,\,\ref{fig:01} (second panel).
To speed up the evaluation of Chebyshev moments, we use a domain-decomposition
technique as implemented in the open-source code \texttt{KITE }\citep{joaoKITEHighperformanceAccurate2020a}.
The proposed approach has two key features: (i) it is sufficiently
powerful to handle systems with millions of sites, thus bypassing
finite-size effects that severely restricted the accessible coupling
strengths in previous studies; and (ii) it provides flexibility to
treat disorder to different levels of accuracy, ranging from a self-consistent
effective-medium-type approximation all the way to a numerically exact
treatment. This offers the possibility to capture important features
of dirty superconductors missed by the commonly employed self-consistent
$T$-matrix approximation, including multiple-impurity quantum interference
phenomena and inhomogeneous pairing patterns.

The mean-field Hamiltonian reads as $\hat{H}=\hat{H}_{0}+\hat{H}_{P}$,
where $\hat{H}_{0}$ is the single-particle Hamiltonian of disordered
graphene {[}Eq. (\ref{eq:graphene}){]} and

\begin{eqnarray}
\hat{H}_{P} & = & E_{0}+g_{0}\sum_{i}\Delta_{0,i}\left(a_{i,\uparrow}^{\dagger}a_{i,\downarrow}^{\dagger}+b_{i,\uparrow}^{\dagger}b_{i,\downarrow}^{\dagger}\right)+\textrm{h.c.}\nonumber \\
 &  & +g_{1}\sum_{\left\langle ij\right\rangle }\Delta_{1,ij}\left(a_{i,\uparrow}^{\dagger}b_{j,\downarrow}^{\dagger}-a_{i,\uparrow}^{\dagger}b_{j,\downarrow}^{\dagger}\right)+\textrm{h.c.}\label{eq:HP}
\end{eqnarray}
is the pairing term. In the above expression, $a_{is}^{\dagger}(b_{is}^{\dagger})$
adds an electron with spin $s=\uparrow,\downarrow$ to site $i$ on
sublattice $A$ ($B$) and $g_{0}$ ($g_{1}$) is the onsite (nearest-neighbour)
interaction energy. The superconducting order parameters are $\Delta_{0,i}=\left\langle a_{i\downarrow}a_{i\uparrow}\right\rangle =\left\langle b_{i\downarrow}b_{i\uparrow}\right\rangle $
and $\Delta_{1,ij}=\left\langle a_{i\downarrow}b_{j\uparrow}-a_{i\uparrow}b_{j\downarrow}\right\rangle $
and $E_{0}=-g_{0}\Delta_{0}^{2}-3g_{1}\Delta_{1}^{2}$ is the condensation
energy.

The clean system achieves quantum criticality at zero chemical potential
($\mu=0$) and displays different superconducting phases depending
on the interaction strength. There are regions of $s$-wave, $p$-wave
and mixed symmetry. Away from half filling, every region is of mixed
symmetry and the quasiparticle spectrum is gapped. In this regime,
the order parameters are smooth across the whole diagram \citep{uchoaSuperconductingStatesPure2007a}.

The order parameters are expressed as

\begin{eqnarray}
\Delta_{0,i} & = & \imath\int_{-\Lambda}^{\Lambda}\text{d}\varepsilon\left(1-2f\left(\omega\right)\right)\langle i,\uparrow|\hat{G}\left(\omega\right)|i,\downarrow\rangle\,,\label{eq:1}\\
\Delta_{1,ij} & = & 2\imath\int_{-\Lambda}^{\Lambda}\text{d}\varepsilon\left(1-2f\left(\omega\right)\right)\langle i,\uparrow|\hat{G}\left(\omega\right)|j,\downarrow\rangle\,,\label{eq:2}
\end{eqnarray}
where $f$ is the Fermi function, $G$ is the retarded Green's function
operator and $i$ and $j$ denote nearest neighbours. The exact choice
of cutoff $\Lambda$ depends on the origin of the pairing term. For
conventional phonon-mediated superconductors, this is the Debye energy
$\hbar\omega_{D}$, which restricts the integration to a thin shell
around the Fermi level. On the other hand, unconventional superconductors,
such as plasmon-mediated metal-coated superconducting graphene \citep{uchoaSuperconductingStatesPure2007a},
generally have contributions from several energy regions. For the
sake of simplicity, we let $\Lambda\rightarrow+\infty$, which captures
the whole spectrum and is still able to accurately reproduce the clean
phase diagram. In the simulations to be presented below, a set of
random sites belonging to either sublattice are selected as the impurity
sites, with a uniform concentration $c=5\%$. Around each impurity
site, the nearest-neighbor hoppings are weakened by an amount $\Delta t$.

Spatial dependence of order parameters---For a clean system, Eqs.~(\ref{eq:1})--(\ref{eq:2})
are easily diagonalizable, yielding a set of two self-consistent equations.
When disorder is introduced, there will be four coupled self-consistent
equations for each lattice site, which severely limits the system
sizes accessible to exact diagonalization. Traditionally, dirty superconductors
have been addressed by means of $T$-matrix and coherent potential
approximation (CPA) schemes \citep{PhysRevLett.77.1849,moradianSuperconductingAlloysWeak2000}.
Here instead, we implement a different strategy that will allow us
to keep the complexity to a minimum. Specifically, we restrict the
order parameters to a constant uniform value in the bulk of the superconductor,
while allowing them to vary in a circle of radius $z$ around each
impurity {[}Fig. \ref{fig04} (a){]}. Since we expect the behavior
of the order parameters to be similar in the vicinity of each impurity,
we further restrict each order parameter to follow an identical spatial
profile $\Delta_{n}=f_{n}(\mathbf{r})$ ($n=0,1$) inside every circle
and then find the function $f_{n}(\mathbf{r})$ which satisfies the
self-consistent equations within these restrictions. This function
is then computed using a stochastic evaluation of the matrix elements
by means of the spectral approach described earlier (further details
are given in Appendix \ref{Appendix_F_superconductor}). This approach
is best suited for dilute point defects, where the regions rarely
overlap. The choice of radius $z$ regulates the approximation. A
larger $z$ allows us to capture more of the spatial dependency of
the order parameters, at the cost of increased running time. The positions
of the impurities are chosen randomly such that no two circles overlap.
The spatial modulation of $\Delta$ is expected to decay very rapidly
\citep{laukeFriedelOscillationsMajorana2018}, so we choose $z$ to
be around three lattice spacings.

Results---Figure \ref{fig04}(b) discloses a rich array of behaviors
for different regions in the $g_{0},g_{1}$ plane. Most of this diagram
for doped graphene can be understood within a virtual crystal approximation,
as if the only effect of disorder is to renormalize the energy scales
of the problem. The effective hopping is given by $t^{\prime}=t+(c/z)\Delta t$,
with $z=3$ the coordination number of the honeycomb lattice. Since
$\Delta t$ is negative, the effective value of the interaction constants
would increase to $g_{0}^{\prime}=g_{0}t/\left(t+c\Delta t/z\right)$
(likewise for $g_{1}$). Such a simple description is able to satisfactorily
explain the behavior seen in regions of increased p-wave and s-wave
order parameters (see Appendix \ref{Appendix_F_SC - 3.Renormalization}).
However, there are two prominent features that cannot be captured
by this heuristic argument. First, the threshold for superconductivity
is substantially reduced when disorder is introduced, a behaviour
typical of the presence of superconducting islands. Secondly, there
is a line in the phase diagram across which the order parameters suffer
an abrupt change that signals the onset of a crossover driven by disorder
{[}white dashed line in Fig. \ref{fig04}(b){]}. To better understand
this transition, we plot in Fig. \ref{fig04}(c) the order parameters
for fixed $g_{1}=-1.7t$ as a function of $g_{0}$ {[}horizontal gray
line in Fig. \ref{fig04}(b){]} for several values of the concentration.
This is a regime of mixed symmetry, since both order parameters are
nonzero. The discontinuity persists even at low (1\%) concentrations
of impurities, but the value of $g_{0}$ for which it happens increases
with increasing concentration. We also plot two matrix elements of
the self-energy as a function of energy at this value of $g_{1}$
for four different values of $g_{0}$ near the transition (Fig. \ref{fig04}d).
The discontinuity in the order parameters is also reflected in these
matrix elements. On both sides of the transition, we see the presence
of a gap due to the finite value of both order parameters, but only
the curves at the right of the transition have a van-Hove-like singularity.
The existence of all off-diagonal matrix elements indicates that this
disorder correlates different sublattices and spins, through the induced
spatial inhomogeneity of the order parameters around the impurities.
The self-energy can therefore provide valuable information about the
local density of states around impurities through its connection to
the LDOS \citep{liImpurityEffectProbe2021}. Since these superconducting
phases may be mediated by plasmons in a proximitized metal layer,
the corresponding coupling parameters may also be controlled by changing
the plasmonic properties of the metal or the distance between the
metal and the graphene sheet. This opens up the possibility of sweeping
the coupling parameters across the discontinuity line to look for
the crossover, which may be identified experimentally through spectroscopic
studies around impurities.

\section{Conclusion}

We introduced a real-space numerical framework that gives access to
the full wavevector and frequency dependence of the quasiparticle
self-energy in arbitrarily complex disordered tight-binding models.
For this purpose, we employed an exact Chebyshev decomposition of
lattice Green's functions (Eq. \ref{eq:CPGF}) that provides full
control over the resolution of the computation. The method was applied
to 3 distinct problems (i.e., quantum criticality driven by chiral
disorder in the honeycomb lattice, impurity resonances in a spin--orbit
coupled ferromagnet and disorder-enhanced superconductivity in monolayer
graphene), in unprecedented large systems, revealing a rich array
of nonperturbative effects that challenge the standard perturbative
picture of disordered systems.

For honeycomb lattices with a dilute concentration of vacancy defects,
we uncovered nonzero off-diagonal components in the self-energy as
well as a strong momentum dependency near the Gade singularity at
zero energy. These previously inaccessible effects are ubiquitous
even in the weak disorder limit, which shows that the quantum interference
of electronic waves scattered by multiple defects plays a much deeper
role than previously thought. A striking result emerges in the long
wavelength limit: the imaginary disorder self-energy at the Dirac
($K$) points vanishes $\textrm{Im}\,\hat{\Sigma}_{\textrm{dis}}(\mathbf{K},0)\simeq0$,
within spectral resolution and accuracy, for any vacancy concentration.
This suggests that the puzzling universal metallic conductivity $\sigma_{\textrm{Gade}}=(4/\pi)e^{2}/h$
previously seen in large-scale quantum transport simulations of defected
graphene systems \citep{ferreiraCriticalDelocalizationChiral2015}
is a fundamental property of chiral-symmetry-protected zero energy
modes with exceedingly large quasiparticle lifetime.

Secondly, our real-space self-energy framework is applied to 2D spin-polarized
electron gases formed in SrRuO$_{3}$/SrTiO$_{3}$ superlattices and
compared against diagrammatic resummation schemes, including the $T$-matrix
for vacancies and the self-consistent Born approximation for uncorrelated
on-site disorder. Both are found to be in excellent agreement with
the numerically exact self-energy at small disorder strength/impurity
concentration, though they diverge significantly away from this regime.
The most distinct effect of a large concentration of vacancies is
the appearance of peaks in the self-energy whose height has a nonperturbative
dependence on the concentration, which we attribute to scattering
resonances from impurity complexes. A rich matrix structure of the
self-energy is borne out by our study, illustrating how correlations
in the disorder potential can manifest as off-diagonal matrix elements
in the self-energy.

Finally, we applied the Chebyshev polynomial Green's function machinery
to calculate the order parameters of superconducting monolayer graphene
in the presence of dilute bond disorder. We found that for some regions
of the phase diagram it is possible to enhance bulk s-wave, p-wave
or both kinds of superconductivity by adjusting the amplitude of local
fluctuations in the hopping parameters. For this purpose, a variation
of the Chebyshev-Bogoliubov-de Gennes method was used to enable self-consistent
simulations of systems with millions of atomic orbitals. These results
open up the intriguing possibility of tailoring superconducting phases
in twisted bilayer graphene, a topic that is currently of much interest.

We briefly comment on possible extensions of the real-space spectral
framework for the disorder self-energy that we introduced in this
work. For conciseness, we restricted ourselves to orthogonal local
basis sets, but this requirement can be easily relaxed at the cost
of introducing an overlap matrix in the calculation of Chebyshev moments
{[}viz. Eq. (\ref{eq:spectral_operator}) and discussion therein{]}.
The use of a nonorthogonal representation of the orbitals would open
doors to accurate studies of disorder effects in complex problems
and materials \citep{Cohen94,Mehl96}. The method can also be easily
extended to evaluate the self-energy resulting from other types of
disorder such as local structures of defects, random hoppings and
correlated disorder. Another interesting question for future study
is whether the framework introduced here could shed new light on the
behavior of mesoscopic systems without self-averaging properties.

\emph{Code availability statement}.---The codes and scripts used
in this study are available from the authors upon reasonable request. 

\emph{Acknowledgements}.---This project was undertaken on the Viking
cluster at the University of York. S.M.J. is supported by Fundação
para a Ciência e Tecnologia (FCT) under the grant no. PD/BD/142798/2018.
S.M.J. and J.M.V.P.L. acknowledge financial support from the FCT,
COMPETE 2020 programme in FEDER component (European Union), through
projects POCI-01-0145-FEDER028887 and UID/FIS/04650/2013, and FCT
through national funds, co-financed by COMPETE-FEDER (grant no. M-ERANET2/0002/2016
-- UltraGraf) under the Partnership Agreement PT2020. A.F. acknowledges
the financial support from the Royal Society through a Royal Society
University Research Fellowship (Grant No. URF\textbackslash R\textbackslash 191021).
We would like to thank J. M. B. Lopes dos Santos, B. Uchoa, J. P.
dos Santos Pires and D. T. S. Perkins for fruitful discussions, and
D. T. S. Perkins for proofreading the final version of the manuscript. 

\bibliographystyle{apsrev4-1}
\bibliography{references}

\pagebreak{}

\section{Appendix}

\subsection{Self averaging property \label{Appendix_A_self_averaging}}

The self-averaging behavior of the disorder self-energy yields highly
converged results in a computationally efficient manner and so must
be justified. While a rigorous general proof is beyond the scope of
this paper, we show below that under some rather general assumptions,
the matrix elements $G_{\alpha\beta}(\mathbf{k},\omega)=\langle\alpha,\mathbf{k}|\hat{G}(\omega)|\beta,\mathbf{k}\rangle$
(and hence the quasiparticle self-energy) satisfy the self-averaging
lemma
\[
\frac{\langle[\textrm{Im}\,G_{\alpha\beta}(\mathbf{k},\omega)]^{2}\rangle-\langle\textrm{Im}\,G_{\alpha\beta}(\mathbf{k},\omega)\rangle^{2}}{\langle\textrm{Im}\,G_{\alpha\beta}(\mathbf{k},\omega)\rangle^{2}}\propto\frac{1}{D}\,,
\]
where $\langle...\rangle$ indicates disorder (configurational) averaging
and $D$ is the Hilbert space dimension of the lattice model that
scales with system volume (a similar expression holds for the real
part of the matrix elements). To simplify the discussion, we specialize
to single-orbital models and thus omit the orbital index $\alpha,\beta$
hereafter. Let $\xi_{\mathbf{k}}=\left\langle \mathbf{k}\right|\hat{G}\left|\mathbf{k}\right\rangle $
denote the matrix element that will be used to determine the self-energy.
It is implied that $\hat{G}=(\omega-\hat{H}+i\eta)^{-1}$ with $\hat{H}=\hat{H}_{0}+\hat{V}_{\textrm{dis}}$.
Specifically, we want to show that the imaginary parts of $\xi_{\mathbf{k}}$
display self-averaging behavior, that is $\text{var}\,\xi_{\mathbf{k}}\equiv\langle(\text{Im}\,\xi_{\mathbf{k}})^{2}\rangle-\langle\text{Im}\,\xi_{\mathbf{k}}\rangle^{2}\propto D^{-1}$.
The argument is identical for the real part of $\xi_{\mathbf{k}}$.
We consider two common classes of problems for lattice models defined
with an arbitrary number of spatial dimensions: (i) systems characterized
by perturbative (weak) disorder effects; and (ii) systems possessing
exponentially localized single-particle states in their spectrum.
Finally, we provide numerical evidence of our claim.

\subsubsection{Weak disorder}

If the diagrammatic expansion of the Green's function is convergent,
then we can use an expansion in powers of $\hat{V}\equiv\hat{V}_{\textrm{dis}}$,
the disorder potential:
\[
\text{Im}\xi_{\mathbf{k}}=\sum_{n=0}^{\infty}\text{Im}\left\langle \mathbf{k}\right|\hat{G}_{0}\left(\hat{V}\hat{G}_{0}\right)^{n}\left|\mathbf{k}\right\rangle 
\]
to evaluate the disorder average $\langle\text{Im}\,\xi_{\mathbf{k}}\,\text{Im}\,\xi_{\mathbf{k}}\rangle$,
keeping in mind that the term $\langle\text{Im}\,\xi_{\mathbf{k}}\rangle\langle\text{Im}\,\xi_{\mathbf{k}}\rangle$
will remove all the terms in the diagrammatic expansion which do not
connect both Green's functions. Defining $G_{0}^{\mathbf{k}}=\left\langle \mathbf{k}\right|\hat{G}_{0}\left|\mathbf{k}\right\rangle $
for convenience and using $\left.\left\langle \mathbf{R}\right|\mathbf{k}\right\rangle =D^{-1/2}e^{i\mathbf{R}\cdot\mathbf{k}}$
to express the disorder potential in real space, we obtain

\begin{eqnarray*}
\\
\text{Im}\,\xi_{\mathbf{k}} & = & \text{Im}G_{0}^{\mathbf{k}}+\text{Im}G_{0}^{\mathbf{k}}\left(\sum_{\mathbf{R}}\frac{1}{D}V_{\mathbf{R}}\right)G_{0}^{\mathbf{k}}+\\
 & + & \text{Im}G_{0}^{\mathbf{k}}\sum_{\mathbf{q}}\sum_{\mathbf{R}\mathbf{R}^{\prime}}\frac{1}{D^{2}}e^{i(\mathbf{R}-\mathbf{R}^{\prime})\cdot(\mathbf{q}-\mathbf{k})}V_{\mathbf{R}}G_{0}^{\mathbf{q}}V_{\mathbf{R}^{\prime}}G_{0}^{\mathbf{k}}\\
 & + & \cdots
\end{eqnarray*}

We get a factor of $1/D$ from every disorder insertion $V_{\mathbf{R}}$
and also a factor of $D$ due to the sum over $\mathbf{R}$. We assume
that $V_{\mathbf{R}}$ is an uncorrelated disorder potential with
Gaussian statistics, i.e. $\langle V_{\mathbf{R}}\rangle=0$, $\langle V_{\mathbf{R}}V_{\mathbf{R}^{\prime}}\rangle\propto\delta_{\mathbf{R}\mathbf{R}^{\prime}}$,
$\langle V_{\mathbf{R}}V_{\mathbf{R}^{\prime}}V_{\mathbf{R}^{\prime\prime}}\rangle=0$,
etc. As explained below this assumption is not essential, but it aids
in substantially simplifying the analysis. The configurational average
introduces correlations between the disorder insertions as Kronecker
deltas $\delta_{\mathbf{R}\mathbf{R}^{\prime}}$ between different
positions. Each $\delta_{\mathbf{R}\mathbf{R}^{\prime}}$ effectively
contributes with an additional factor of $1/D$. Lastly, each loop
in the diagrams (representing integrations over internal momenta)
contributes with another factor of $D$. 

\begin{figure}
\begin{centering}
\includegraphics[scale=0.7]{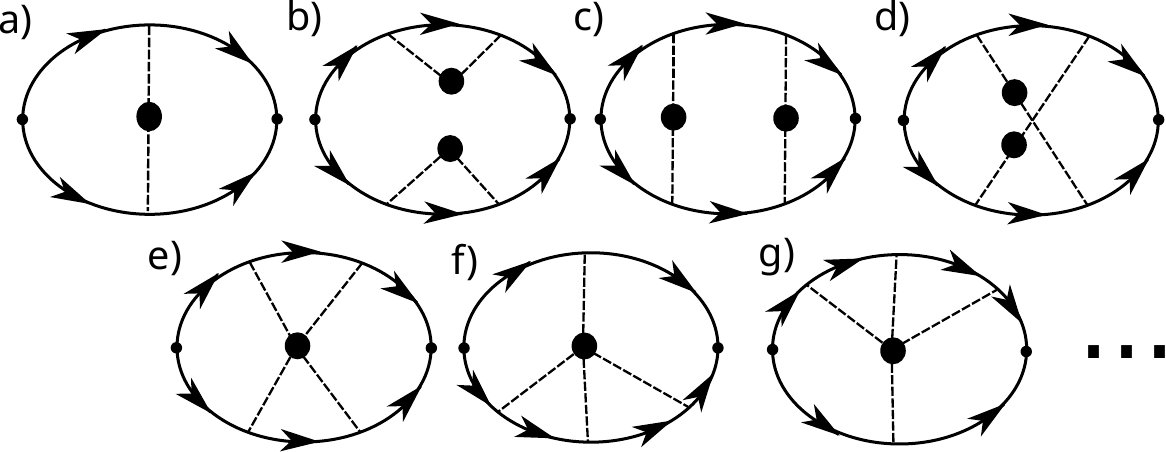}
\par\end{centering}
\caption{\label{fig_supp_01}Feynman diagrams that contribute to the variance
of $\text{Im}\,G_{\mathbf{k}}$ up to fourth order in $\hat{V}_{\text{dis}}$.}

\end{figure}

Figure \ref{fig_supp_01} shows the diagrams that contribute to the
variance up to fourth order in $\hat{V}_{\text{dis}}$. Counting all
the powers of $D$, one can check that each term is associated with
a factor of $1/D$ except for diagram (b). Instead, this diagram is
proportional to $(D-1)/D$, but the constant term gets cancelled precisely
by $\langle\text{Im}\xi_{\mathbf{k}}\rangle\langle\text{Im}\xi_{\mathbf{k}}\rangle$
and what is left is again proportional to $1/D$. At higher orders
in $\hat{V}_{\text{dis}}$, similar arguments can be made. If the
upper branch of the diagrams is not connected to the lower branch,
then it will get almost completely cancelled by $\langle\text{Im}\xi_{\mathbf{k}}\rangle^{2}$,
leaving only the $1/D$ contribution. If both branches are connected,
the number of loops is not large enough to destroy the $1/D$ dependency.

While we have only strictly presented our argument for uncorrelated
disorder, we argue that a generalization to correlated disorder should
also be possible provided that the correlation length is finite. In
such a scenario, averaging over disorder would introduce asymptotically
decreasing functions of the distance between $\mathbf{R}$ and $\mathbf{R}^{\prime}$
in lieu of Kronecker deltas. In any case, a sum over the position
(which would contribute with a factor of $D$ as noted above) now
contributes with a factor of order unity, effectively having the same
effect as the Kronecker delta for the purposes of self averaging.

\subsubsection{Localized states}

Next, we analyze an important class of problems where diagrammatic
methods break down \citep{PhysRevB.22.4666}: strongly disordered
systems with localized states in their spectrum. We assume that the
value of $\omega$ is such that all the states in an energy window
$\eta$ around $\omega$ are localized, with a maximum localization
length of $\zeta$. We begin by expressing $\xi_{\mathbf{k}}$ in
terms of the eigenstates with energies $\{\varepsilon_{\alpha}\}$
resolved in space

\[
\xi_{\mathbf{k}}=\frac{1}{D}\sum_{\mathbf{R}\mathbf{R}^{\prime}\alpha}e^{i\mathbf{k}\cdot\text{\ensuremath{\left(\ensuremath{\mathbf{R}^{\prime}}-\mathbf{R}\right)}}}\frac{\left.\left\langle \mathbf{R}\right|\alpha\right\rangle \left.\left\langle \alpha\right|\mathbf{R}^{\prime}\right\rangle }{\omega+i\eta-\varepsilon_{\alpha}}=\frac{1}{D}\sum_{\mathbf{R}}g_{\mathbf{R},\mathbf{k}}
\]
where $g_{\mathbf{R},\mathbf{k}}$ represents the contribution to
$\xi_{\mathbf{k}}$ from the sites around $\mathbf{R}$. By assumption,
these states are localized, so, for each $\mathbf{R}$, only localized
states with localization center within a distance $2\zeta$ around
$\mathbf{R}$ contribute. Let $S$ be this region. This means that
$g_{\mathbf{R},\mathbf{k}}$ and $g_{\mathbf{R}^{\prime},\mathbf{k}}$
have appreciable correlation only if $\left|\mathbf{R}-\mathbf{R}^{\prime}\right|<2\zeta$.
It is important to note that $g_{\mathbf{R},\mathbf{k}}$ is independent
of the system size, since the percentage of localized states is assumed
to be an intensive property. This is a key assumption of this proof
and fundamentally relies on the existence of a mobility edge. Note
that $g_{\mathbf{R},\mathbf{k}}$ is a random variable with a finite
maximum absolute value because only a finite number $N_{e}$ of elements
contribute to both the sum over $\mathbf{R}^{\prime}$ and the sum
over $\alpha$. Using the triangle inequality, 
\begin{eqnarray*}
\left|g_{\mathbf{R},\mathbf{k}}\right| & \leq & \sum_{\mathbf{R}^{\prime}\alpha}\frac{|\langle\mathbf{R}|\alpha\rangle|\,|\langle\alpha|\mathbf{R}^{\prime}\rangle|}{\left(\omega-\varepsilon_{\alpha}\right)^{2}+\eta^{2}}\\
 & \leq & \eta^{-2}\sum_{\mathbf{R}^{\prime}\alpha}|\langle\mathbf{R}|\alpha\rangle|\,|\langle\alpha|\mathbf{R}^{\prime}\rangle|.
\end{eqnarray*}
For both sums, $N_{e}$ is the number of degrees of freedom inside
a $d$-dimensional sphere of radius $2\zeta$. Therefore, the sum
$D^{-1}\sum_{\mathbf{R}}g_{\mathbf{R},\mathbf{k}}$ can be seen as
a sum of bounded random variables which are only correlated within
a distance $\left|\mathbf{r}\right|<2\zeta$ of one another. Thus
$\xi_{\mathbf{k}}$ follows the central limit theorem and so $\text{var}\,\xi_{\mathbf{k}}\sim D^{-1/2}$,
hence proving the self-averaging property. We note the only assumptions
made in this derivation were the locality of the localized wave functions
and that only localized wave functions have relevant spectral weight
in $G_{\mathbf{k}}$.

\subsubsection{Numerical demonstration}

\begin{figure}
\begin{centering}
\includegraphics[width=0.9\columnwidth]{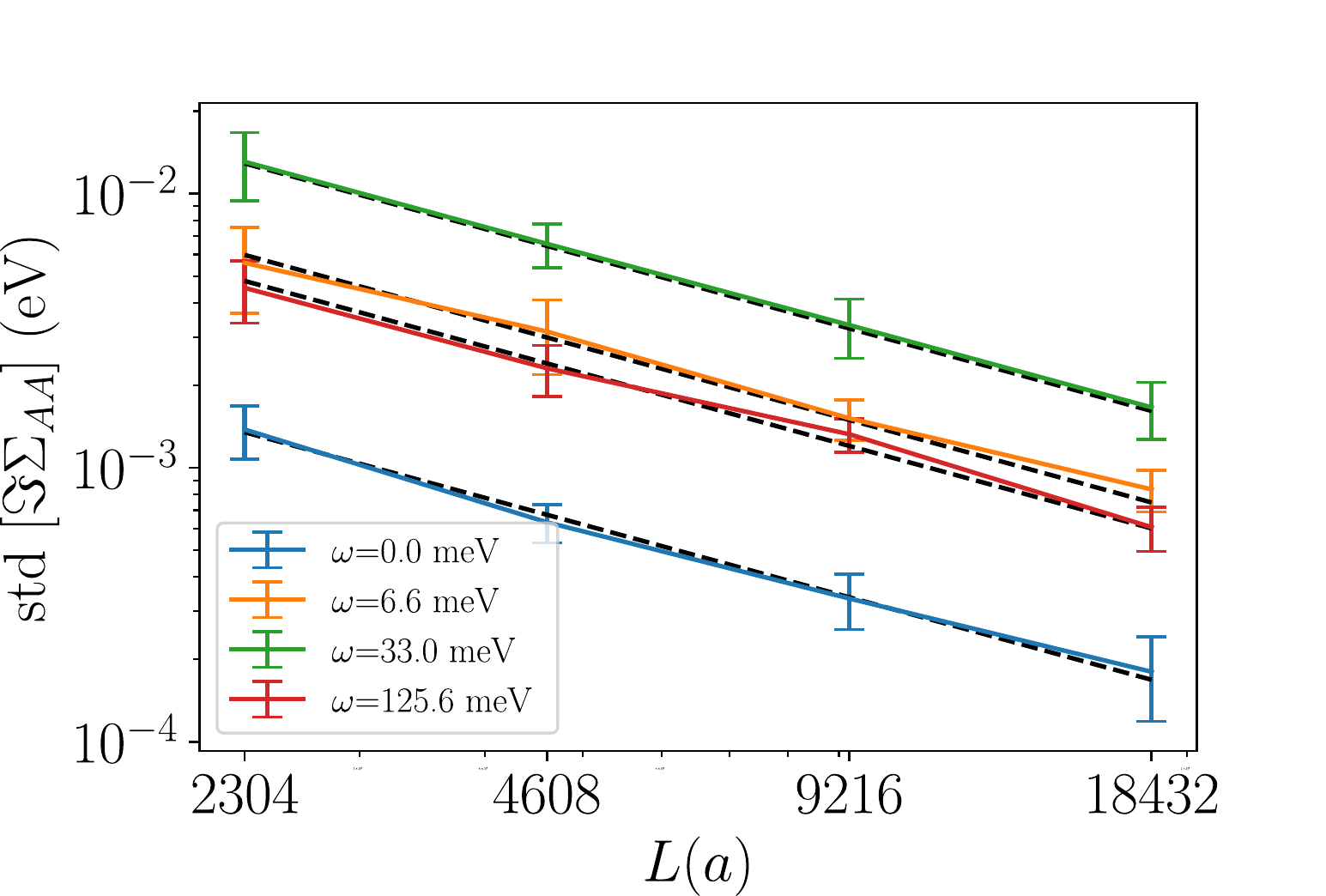}
\par\end{centering}
\caption{\label{fig_supp_02}System size dependence of the standard deviation
of the the AA component of the disorder self-energy matrix of graphene
at the Dirac point $\mathbf{k}=\mathbf{K}$ for selected energies.
Here we used a concentration of $0.3\%$ of vacancies and a broadening
of 1 meV in the spectral evaluation of the Green's function. The same
realization of disorder was used in the calculation of the matrix
elements of the full Green's function $G^{\eta}(\mathbf{k},\omega)$
needed to evaluate the self-energy via Eq. (\ref{eq:se_formula}).
This prescription is highly effective in improving the convergence
of the ensemble average. Error bars are shown.}
\end{figure}

Next, we demonstrate the self-averaging behavior numerically for the
nontrivial example of a graphene system hosting a Gade singularity
at the band center ($\omega=0$) generated by dilute point defects.
To this end, we calculated the self-energy at the $\mathbf{k}=\mathbf{K}$
point of very large lattices ($D=2L^{2}=10616832,42467328,169869312$
and $679477248$) for 100 realizations of disorder. This allowed us
to obtain the standard deviation of this stochastic quantity (colored
curves in Fig. \ref{fig_supp_02}), as a function of the linear system
size. If self-averaging is taking place, then the standard deviation
should decrease as $\sim L^{-1}$, which is exactly what we obtain.
The black dashed curves in Fig. \ref{fig_supp_02} indeed have slope
of $-1$ in a log--log scale, which indicates the correct scaling
law.

\subsection{Multi-scale domain decomposition and other computational details
\label{Appendix_B_domain_decomp}}

Computations for this work were carried out with the open-source \texttt{KITE}
code \citep{joaoKITEHighperformanceAccurate2020a}. Periodic boundary
conditions were employed in all calculations. \texttt{KITE} implements
an efficient decomposition of the exact Green's function in terms
of Chebyshev polynomials, which was used both for the calculation
of the self-energy in all case studies as well as the local Green's
function in the superconductor problem in Sec. \ref{subsec:Example_3_pwave}.
The real-space Green's functions are evaluated using the CPGF approach
{[}Eq. (\ref{eq:CPGF}){]} and the calculation of $\langle\alpha,\mathbf{k}|G(\omega)|\beta,\mathbf{k}\rangle$
(or $\langle\alpha,\mathbf{R}|G(\omega)|\beta,\mathbf{R}^{\prime}\rangle$
for the superconductor problem) relies entirely on evaluating the
Chebyshev moments $\langle\alpha,\mathbf{k}|T_{n}(\hat{H})|\beta,\mathbf{k}\rangle$.
The recursive nature of our spectral approach means that the complexity
of the calculation of this matrix element scales linearly with the
number of polynomials $M$.

Every object in the spectral approach is expressed in real space,
exploiting the sparseness of the Hamiltonian matrix to improve the
parallelization performance during the computation of the Chebyshev
moments. In addition to being sparse, this Hamiltonian $\hat{H}$
typically only connects sites that are close by neighbors. In the
process of the matrix-vector multiplication $|v^{\prime}\rangle=\hat{H}|v\rangle$,
\texttt{KITE} divides the vector $|v\rangle$ into equally-sized domains,
which get assigned to different processing cores, and further subdivides
these domains into tiles. Within each core, this multiplication is
completed within each tile first, before moving on to the next tile
to minimize cache misses when bringing the tile from memory. The size
of the tile is adjusted according to the processor's cache size to
maximize performance. This is at the core of the parallelization scheme
in \texttt{KITE} \citep{joaoKITEHighperformanceAccurate2020a}. Each
processor performs the real-space matrix multiplication within its
assigned domain, but it requires information about the other neighboring
domains to correctly perform the multiplication around the borders.
This is mitigated by keeping a copy of the neighboring domains' borders
stored in memory for each processor. After the matrix-vector product
has been calculated, the stored copy of the borders is updated before
proceeding to the next iteration. This step is not parallelizable,
but it scales with the area of the lattice rather than the volume,
so the algorithm becomes more efficient with increasing lattice size.

In Sec. \ref{subsec:A}, the $\mathbf{k}$ points chosen for the calculation
of the self-energy belong to the (finite discrete) first Brillouin
zone. Since the $K$ point of the honeycomb lattice may be expressed
as $\left(\mathbf{b}_{1}-\mathbf{b}_{2}\right)/3$, in terms of the
reciprocal lattice primitive vectors, the linear system sizes used
were limited to multiples of $3$. Failing to do so may induce noticeable
errors due to the overlap with other momenta coming from the decomposition
of $\mathbf{k}$ in terms of vectors belonging to the first Brillouin
zone.
\begin{figure}
\begin{centering}
\includegraphics[width=0.9\columnwidth]{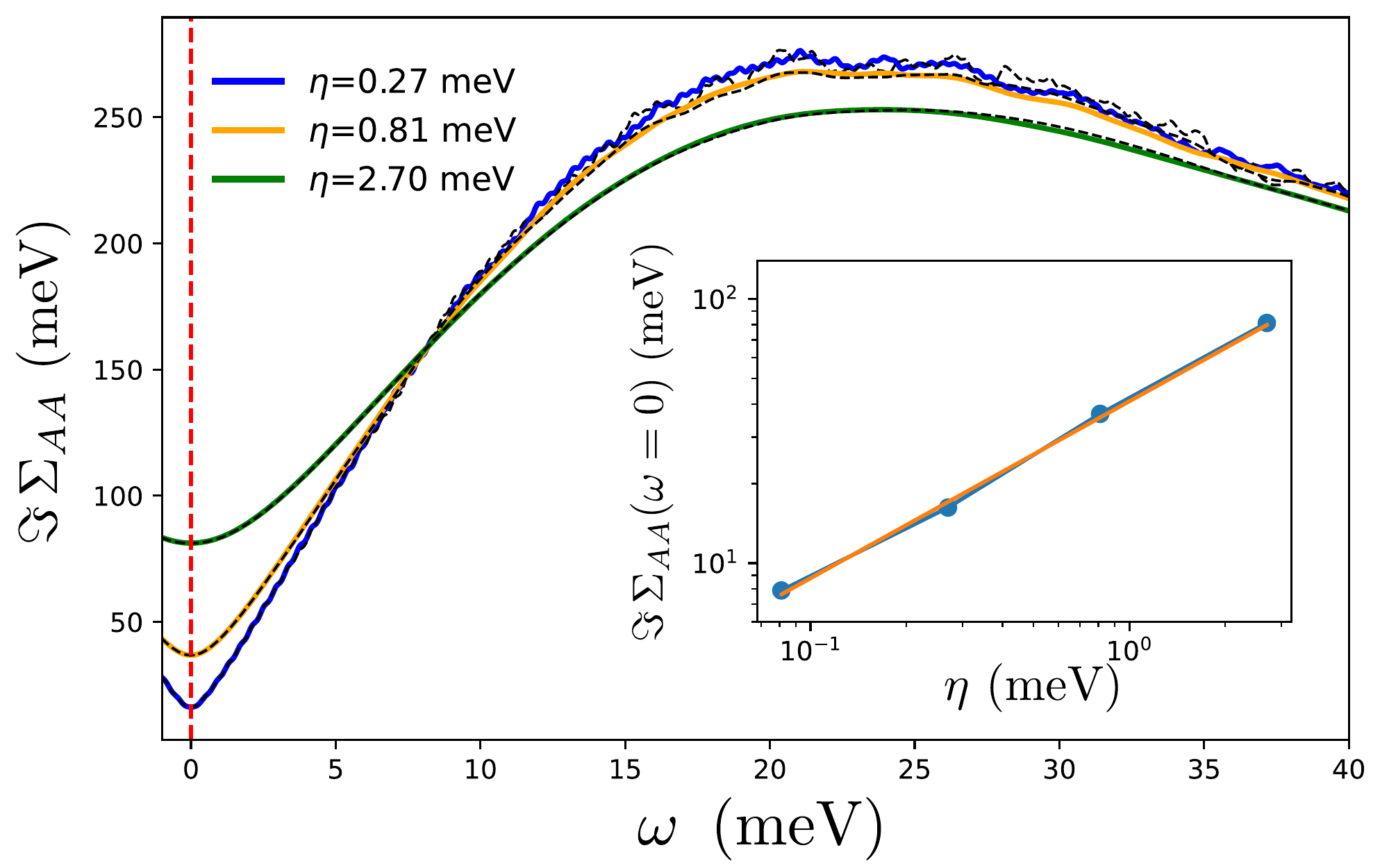}
\par\end{centering}
\caption{\label{fig_supp_03}Convergence study of the energy-dependence of
the AA component of the self-energy matrix at the Dirac point $\mathbf{k}=\mathbf{K}$.\textcolor{blue}{{}
}\textcolor{black}{Calculations are carried out on large lattices
with linear dimen}sions $L_{x}=L_{y}=36864a$ ($D\simeq2.7\times10^{9}$
orbitals). Each colored curve represents a different broadening $\eta$
and the superimposed dashed black curves are for a smaller system
with $L_{x}=L_{y}=18432a$. The inset shows the self-energy at zero
energy as a function of $\eta$. Here we used a vacancy concentration
of $0.3\%$.}
\end{figure}

\subsection{Spectral convergence\label{Appendix_C_spectral_conv}}

Convergence against several factors has been carefully assessed in
all problems studied in this work, specifically:
\begin{enumerate}
\item The energy resolution $\eta$ has to be as small as possible to accurately
capture the singular nature of the Green's functions;
\item For any $\eta$, convergence of the Chebyshev series (i.e., the choice
of truncation order $M-1$) to the desired accuracy needs to be established
carefully;
\item Linear dimensions need to be large enough for the mean-level spacing
to be suitably small compared to the target resolution, $\delta\varepsilon\lesssim\eta$,
and;
\item There may be strong fluctuations arising from specific realizations
of disorder. A larger system size helps removing such artifacts.
\end{enumerate}
Here we address points 1--4 with the help of Fig. \ref{fig_supp_03},
where we show the disorder self-energy of graphene calculated for
a selected concentration of vacancies (0.3\%), two different system
sizes and several broadening factors $\eta$.
\begin{enumerate}
\item Due to the singular nature of this problem and the very fine resolutions
required, full convergence with $\eta$ is challenging to achieve,
particularly at the band center. Sufficiently far from $\omega=0$,
the orange curve in Fig. \ref{fig_supp_03} appears to be converged.
The inset shows the self-energy dependence as a function of $\eta$
for states with $\omega=0$. This curve is well fitted by $\text{Im}\,\Sigma_{AA}(\mathbf{K},0)=\eta^{2/3}$
(orange line) which at $\eta=0$ extrapolates to $\text{Im}\,\Sigma_{AA}(\mathbf{K},0)=0$.
\item The curves no longer change when we increase further the number of
polynomials, indicating that the Chebyshev series used to calculate
the Green's functions is fully converged. The maximum number of polynomials
was $131072$.
\item In Fig. \ref{fig_supp_03}, we represented $\Sigma_{AA}(\mathbf{K},\omega)$
calculated with $L_{x}=L_{y}=36864a$, with $a$ the lattice spacing,
for three different values of $\eta$ in different colors. Each of
these colors has a black dashed curve superimposed, which is the same
calculation done with $L_{x}=L_{y}=18432a$. Close to zero energy,
there are no appreciable differences upon changing the system size.
This indicates that the mean-level spacing is very small near the
band center (due to the large density of states in this region), thus
leading to smoother, fully converged curves. Away from this region,
the discreteness of the spectrum starts to become visible at finer
resolutions and larger system sizes mitigate this effect.
\item The statistical fluctuations due to the disorder realizations are
very small, of the order of 2 meV when $L_{x}=L_{y}=18432a$ and therefore
do not influence our results (see previous section).
\end{enumerate}

\subsection{SRO \label{Appendix_D_SRO}}

\subsubsection{Multi-orbital model\label{Appendix_D_SRO-1.Hamiltonian}}

The Hamiltonian used in Sec. \ref{subsec:Example_2_SRO} consists
of a six-orbital tight-binding model on a square lattice, which can
be divided into four terms:

\[
\hat{H}=\hat{H}_{1}+\hat{H}_{2}+\hat{H}_{3}+\hat{H}_{4}.
\]
Here, 
\begin{eqnarray*}
\hat{H}_{1} & = & \sum_{a,\sigma,\left\langle ij\right\rangle _{x}}t^{a,x}d_{ia\sigma}^{\dagger}d_{ja\sigma}+\sum_{a,\sigma,\left\langle ij\right\rangle _{y}}t^{a,y}d_{ia\sigma}^{\dagger}d_{ja\sigma}
\end{eqnarray*}
represents the nearest-neighbor interactions, with $t^{1,x}=t^{2,y}=t_{2}$,
$t^{2,x}=t^{3,x}=t^{1,y}=t^{3,y}=t_{1}$. The operator $d_{ia\sigma}^{\dagger}$
creates an electron in site $i$ orbital $a$ ($yz=1$, $xz=2$, $xy=3$)
and spin projection $\sigma$. The second term
\[
\hat{H}_{2}=\sum_{a,b,\sigma,\left\langle \left\langle i,j\right\rangle \right\rangle }f_{ij}^{ab}d_{ia\sigma}^{\dagger}d_{jb\sigma}+\sum_{a,\sigma,\left\langle \left\langle i,j\right\rangle \right\rangle }g^{a}d_{ia\sigma}^{\dagger}d_{ja\sigma}
\]
is the second-nearest-neighbor contribution to the Hamiltonian with
$g^{1}=g^{2}=t_{3}$, $g^{3}=t_{4}$ and $f_{ij}^{12}=f_{ij}^{21}=f$
if $i$ and $j$ are along the diagonal and $f_{ij}^{12}=f_{ij}^{21}=-f$
if they are along the anti\-diagonal. The terms $\hat{H}_{3}$ and
$\hat{H_{4}}$ encode the Zeeman interaction and spin--orbit coupling
(SOC), respectively, with the expressions

\begin{eqnarray*}
\hat{H}_{3} & = & -m\sum_{a}\sum_{\sigma,i}\tau_{\sigma\sigma}^{z}d_{ia\sigma}^{\dagger}d_{ia\sigma}\\
\hat{H}_{4} & = & i\lambda\sum_{a,b,c,}\sum_{\sigma,\sigma^{\prime},i}\varepsilon^{abc}\tau_{\sigma\sigma^{\prime}}^{c}d_{ia\sigma}^{\dagger}d_{ib\sigma^{\prime}}
\end{eqnarray*}
where $m$ is the amplitude of the Zeeman interaction, $\lambda$
is the amplitude of the SOC, $\tau^{i}$ is a Pauli matrix ($\tau^{1}=\sigma_{x}$,
$\tau^{2}=\sigma_{y}$ and $\tau^{3}=\sigma_{z}$) and $\varepsilon^{abc}$
is the Levi-Civita symbol. The SOC term was calculated by evaluating
the matrix elements $\hat{\mathbf{L}}\cdot\hat{\mathbf{S}}$ in the
angular momentum basis with $\ell=2$ restricted to the Cartesian
set (i.e. $\{xy,xz,yz\}$).

\subsubsection{Self-energy: other matrix elements\label{Appendix_D_SRO - 2.Other-matrix-elements}}

\begin{figure}
\begin{centering}
\includegraphics[width=1\columnwidth]{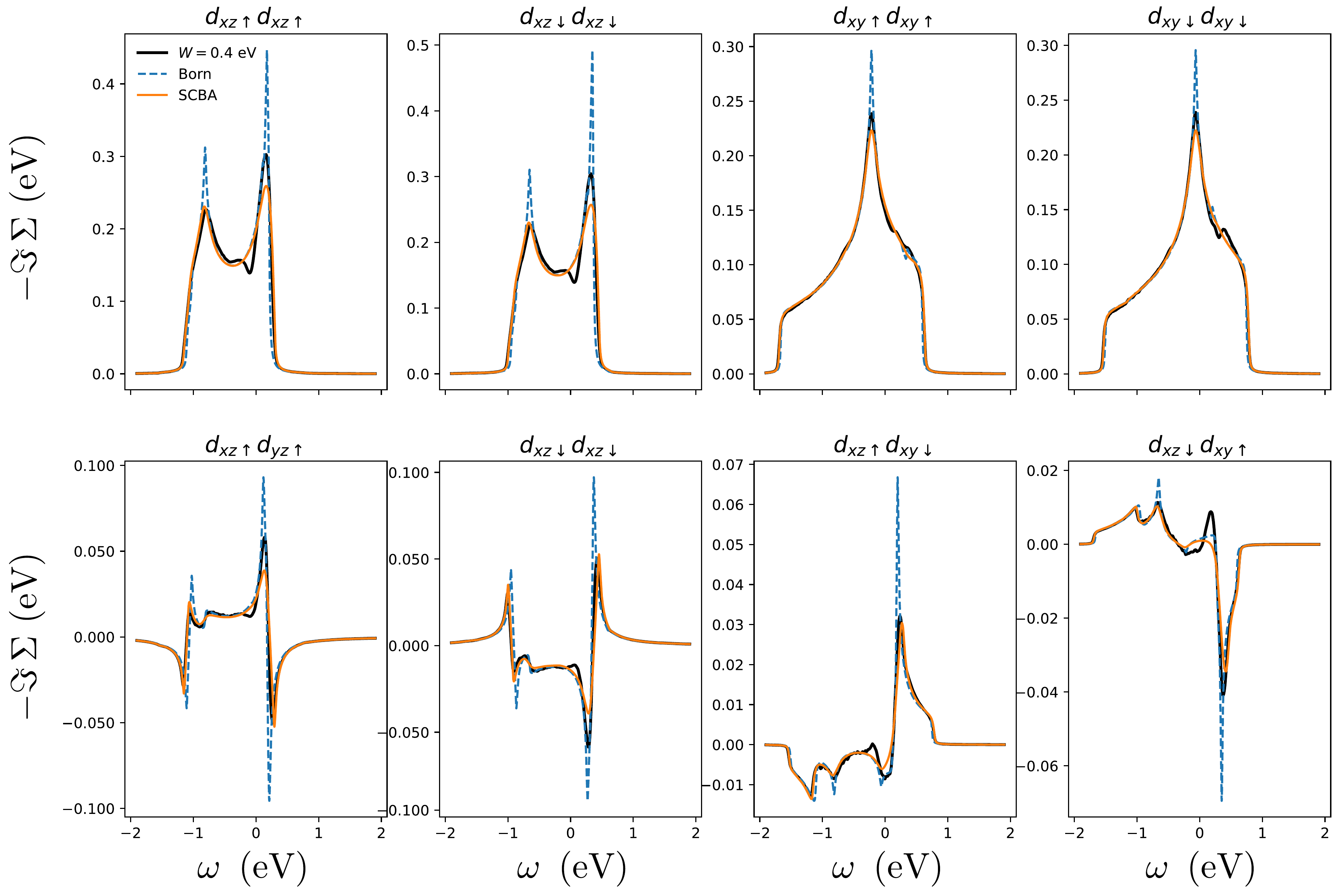}
\par\end{centering}
\caption{\label{fig_supp_04}Imaginary part of all the eight independent matrix
elements of the disorder self-energy calculated for SRO with Anderson
disorder at the $\Gamma$ point. The calculation details are the same
as in section \ref{subsec:Example_2_SRO}.}
\end{figure}

Figure \ref{fig_supp_04} displays the non-zero independent self-energy
matrix elements for the example of SRO with Anderson disorder. The
remaining $10$ nonzero matrix elements of the self-energy are related
to these according to the following relations $\Sigma_{d_{yz}\uparrow,d_{yz}\uparrow}=\Sigma_{d_{xz}\uparrow,d_{xz}\uparrow}$,
$\Sigma_{d_{xz}\uparrow,d_{yz}\uparrow}=-\Sigma_{d_{yz}\uparrow,d_{xz}\uparrow}$,
$\Sigma_{d_{yz}\downarrow,d_{yz}\downarrow}=\Sigma_{d_{xz}\downarrow,d_{xz}\downarrow}$,
$\Sigma_{d_{yz}\downarrow,d_{xz}\downarrow}=-\Sigma_{d_{xz}\downarrow,d_{yz}\downarrow}$,
$\Sigma_{d_{yz}\downarrow,d_{xy}\uparrow}=\Sigma_{d_{xy}\uparrow,d_{yz}\downarrow}=i\Sigma_{d_{xz}\downarrow,d_{xy}\uparrow}=\Sigma_{d_{xy}\uparrow,d_{xz}\downarrow}$,
and $\Sigma_{d_{yz}\uparrow,d_{xy}\downarrow}=\Sigma_{d_{xy}\downarrow,d_{yz}\uparrow}=-i\Sigma_{d_{xz}\uparrow,d_{xy}\downarrow}=i\Sigma_{d_{xy}\downarrow,d_{xz}\uparrow}$.

\subsection{Disorder operator correlations\label{Appendix_E_disorder_correlations}}

In this section we show that the elements appearing in the self-energy
matrix depend on the type of correlations appearing in the disorder
potential. We find the dependency explicitly within the first Born
approximation. In this paper, we use two kinds of disorder: Anderson
disorder and dilute (short-range) disorder. We perform this analysis
for Anderson disorder, but the proof is similar for the dilute case\@.
Consider the following disorder operator $\hat{V}\equiv\hat{V}_{\textrm{dis}}$
in the Hamiltonian. The disorder operator $V$ is diagonal in real
space and is given by

\[
\hat{V}=\sum_{i\alpha}W_{i\alpha}|i,\alpha\rangle\langle i,\alpha|
\]
where $W_{i\alpha}$ is taken from the uniform distribution {[}see
e.g., Eq. (\ref{eq: SRO - distribution Anderson}){]} with width $W$
and mean $0$. Here, $i$ indexes the unit cell and $\alpha$ indexes
the degrees of freedom within the unit cell such as spin, orbital
and/or sublattices. In momentum space,
\[
\hat{V}=\sum_{\boldsymbol{p}\boldsymbol{p}^{\prime}\alpha}V_{\boldsymbol{p}\boldsymbol{p}^{\prime}}^{\alpha\beta}|\boldsymbol{p},\alpha\rangle\langle\boldsymbol{p}^{\prime},\beta|
\]
where 
\[
V_{\boldsymbol{p},\boldsymbol{p}^{\prime}}^{\alpha\beta}=\frac{1}{N}\sum_{j}\exp\left(i\boldsymbol{R}_{j}\cdot\left(\boldsymbol{p}^{\prime}-\boldsymbol{p}\right)\right)W_{j\alpha}\delta_{\alpha\beta}=V_{\boldsymbol{p}-\boldsymbol{p}^{\prime}}^{\alpha\beta}
\]
and $\boldsymbol{R}_{j}$ denotes the position of the $j$-th unit
cell. The average of two disorder operators is

\begin{eqnarray*}
\left\langle V_{\boldsymbol{p}}^{\alpha\beta}V_{\boldsymbol{p}^{\prime}}^{\gamma\delta}\right\rangle  & = & \frac{1}{N^{2}}\left\langle \sum_{j}e^{i\boldsymbol{R}_{j}\cdot\boldsymbol{p}}W_{j\alpha}\delta_{\alpha\beta}\sum_{i}e^{i\boldsymbol{R}_{i}\cdot\boldsymbol{p}^{\prime}}W_{i\gamma}\delta_{\gamma\delta}\right\rangle \\
 & = & \frac{1}{N^{2}}\sum_{ij}e^{i\boldsymbol{R}_{j}\cdot\boldsymbol{p}}e^{i\boldsymbol{R}_{i}\cdot\boldsymbol{p}^{\prime}}\left\langle W_{j\alpha}W_{i\gamma}\right\rangle \delta_{\alpha\beta}\delta_{\gamma\delta}
\end{eqnarray*}
Now we assume that $W_{i\alpha}$ is uncorrelated among different
unit cells, but there might be correlations within any given unit
cell. For example, if the $\alpha$ indices refer to spin up and down,
one might require that $W_{i\uparrow}=W_{i\downarrow}$, which would
impose $\langle W_{j\alpha}W_{i\gamma}\rangle=\delta_{ij}\frac{W^{2}}{12}$
for arbitrary $\alpha$ and $\gamma$. In contrast, if one is dealing
with graphene and the $\alpha$ indices refer to the sublattice, one
might require that the value of $W$ in $A$ be independent of that
in $B$, yielding $\langle W_{j\alpha}W_{i\gamma}\rangle=\delta_{ij}\delta_{\alpha\gamma}\frac{W^{2}}{12}$.
With more generality, let's assume $\langle W_{j\alpha}W_{i\gamma}\rangle=\delta_{ij}M_{\alpha\gamma}$,
where $M$ is a matrix which captures the correlations. Then, 
\begin{eqnarray*}
\left\langle V_{\boldsymbol{p}}^{\alpha\beta}V_{\boldsymbol{p}^{\prime}}^{\gamma\delta}\right\rangle  & = & \frac{1}{N^{2}}\sum_{ij}e^{i\boldsymbol{R}_{j}\cdot\boldsymbol{p}}e^{i\boldsymbol{R}_{i}\cdot\boldsymbol{p}^{\prime}}\delta_{ij}M_{\alpha\gamma}\delta_{\alpha\beta}\delta_{\gamma\delta}\\
 & = & \frac{1}{N^{2}}\sum_{j}e^{i\boldsymbol{R}_{j}\cdot\left(\boldsymbol{p}+\boldsymbol{p}^{\prime}\right)}M_{\alpha\gamma}\delta_{\alpha\beta}\delta_{\gamma\delta}\\
 & = & \frac{1}{N}\delta_{\boldsymbol{p}+\boldsymbol{p}^{\prime}}M_{\alpha\gamma}\delta_{\alpha\beta}\delta_{\gamma\delta}
\end{eqnarray*}
The self-energy is calculated perturbatively by expanding the disorder-averaged
Green's function in a power series of the disorder operator
\[
\Sigma_{\boldsymbol{p}\boldsymbol{p}^{\prime}}^{\alpha\beta}\left(\varepsilon\right)=\left\langle V_{\boldsymbol{p}\boldsymbol{p}^{\prime}}^{\alpha\beta}\right\rangle +\sum_{\boldsymbol{q}\beta\gamma}\left\langle V_{\boldsymbol{p}\boldsymbol{q}}^{\alpha\gamma}G_{\boldsymbol{q}\gamma\delta}^{0}V_{\boldsymbol{q}\boldsymbol{p}^{\prime}}^{\delta\beta}\right\rangle +\cdots
\]
Here, only the 1-point irreducible diagrams are to be kept. In the
Born approximation, we obtain

\begin{eqnarray*}
\Sigma_{\boldsymbol{p}\boldsymbol{p}^{\prime}}^{\alpha\beta}\left(\varepsilon\right) & = & \sum_{\boldsymbol{q}\beta\gamma}\left\langle V_{\boldsymbol{p}-\boldsymbol{q}}^{\alpha\gamma}V_{\boldsymbol{q}-\boldsymbol{p}^{\prime}}^{\delta\beta}\right\rangle G_{\boldsymbol{q}\gamma\delta}^{0}\\
 & = & \sum_{\boldsymbol{q}\beta\gamma}\frac{1}{N}\delta_{\boldsymbol{p}-\boldsymbol{q}+\boldsymbol{q}-\boldsymbol{p}^{\prime}}M_{\alpha\delta}\delta_{\alpha\gamma}\delta_{\delta\beta}G_{\boldsymbol{q}\gamma\delta}^{0}\\
 & = & \left(\frac{1}{N}\sum_{\boldsymbol{q}}G_{\boldsymbol{q}\alpha\beta}^{0}\right)M_{\alpha\beta}\delta_{\boldsymbol{p}-\boldsymbol{p}^{\prime}}
\end{eqnarray*}
and it is clear that the matrix elements appearing in $\Sigma$ depend
on the correlation matrix $M$. If the disorder is completely uncorrelated,
only the diagonal elements will survive. Despite this, higher-order
terms beyond the Born approximation may contribute to the off-diagonal
matrix elements of the self-energy.

\subsection{Superconductor\label{Appendix_F_superconductor}}

\subsubsection{Computational details\label{Appendix_F_SC - 1.method}}

Here we provide additional details on the superconducting order parameter
calculation. The starting point is the Chebyshev-Bogoliubov-de Gennes
formalism \citep{covaciEfficientNumericalApproach2010}, where the
order parameters are obtained from Eqs. \ref{eq:1} and \ref{eq:2}.
In the clean case, the order parameters share the periodicity of the
crystal. In the presence of disorder, we expect the order parameters
to change appreciably in the vicinity of impurities and defects, but
to remain relatively constant otherwise. In some cases, this modulation
around an impurity may extend up to a few dozens of nanometers \citep{kimLongrangeFocusingMagnetic2020},
but in our case, with dilute nonmagnetic impurities in graphene, the
order parameter only changes appreciably on the order of a few unit
cells \citep{laukeFriedelOscillationsMajorana2018}. 

For this task, we propose a new approach to calculate the order parameters
which takes into account most of the spatial modulation while only
requiring a small number of self-consistent equations be solved. Taking
$\Delta_{0,i}$ as an example (an identical procedure is used for
$\Delta_{1,ij}$), we start by defining a circle centered around each
impurity site $\boldsymbol{R}_{n}$ with radius $z$. Outside of these
circles, the order parameter $\Delta_{0,i}$ is restricted to be uniform:
$\Delta_{0,i}=\Delta_{0}^{b}$. Inside every circle, the order parameter
is assumed to behave identically, so for the sites $\boldsymbol{r}_{i}$
such that $\left|\boldsymbol{R}_{n}-\boldsymbol{r}_{i}\right|<z$,
the order parameter satisfies $\Delta_{0}\left(\boldsymbol{R}_{n}+\boldsymbol{r}_{i}\right)=\rho_{0,i}$
for every $\boldsymbol{R}_{n}$ and for some (yet to be determined)
function $\rho_{0,i}$. In each self-consistent step, the value for
the next $\rho_{0,i}$ is defined as an average over identical sites
within each circle $\rho_{0,i}=N_{\textrm{imp}}^{-1}\sum_{n=1}^{N_{\textrm{imp}}}\Delta_{0}\left(\boldsymbol{R}_{n}+\boldsymbol{r}_{i}\right)$,
where $N_{\textrm{imp}}$ is the number of impurities. $\Delta_{0}^{b}$
is calculated by averaging over the remaining sites. With this method,
the number of self-consistent equations that have to be solved scales
with the area of one circle $\sim z^{2}$ instead of the number of
lattice sites. We are able to greatly reduce the number of self-consistent
equations that have to be solved while still capturing most of the
spatial dependency of $\Delta_{0,i}$. The radius of the circles can
be adjusted in order to better reflect the spatial dependency of the
order parameters around the impurities. 
\begin{figure}
\begin{centering}
\includegraphics[width=1\columnwidth]{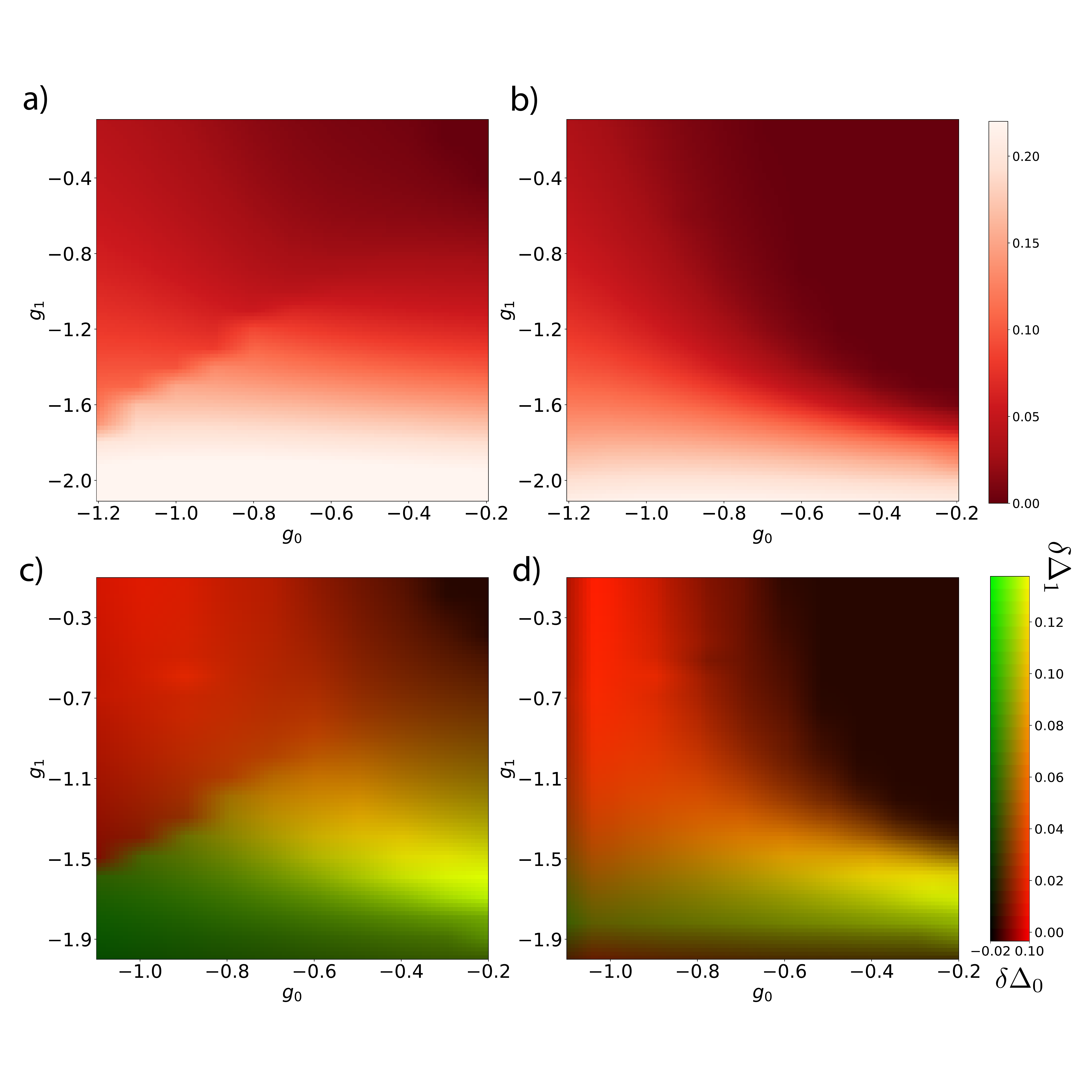}
\par\end{centering}
\caption{Superconducting order parameter $\Delta_{1}$ across the pairing plane
with (a) and without (b) disorder. Change in the superconducting order
parameter due to presence of impurities (c) and renormalization of
the energy scales (d). The color scale in (c) and (d) is the same
as in Fig. \ref{fig04}.}
\label{fig_supp_05}
\end{figure}

Coming back to Eq. \ref{eq:1}, the new self-consistent equation that
has to be solved is

\begin{eqnarray*}
\rho_{0,i} & = & \imath\int_{-\Lambda}^{\Lambda}\text{d}\omega\left(1-2f\left(\omega\right)\right)\times\\
 &  & \times\frac{1}{N_{\textrm{imp}}}\sum_{n=1}^{N_{\textrm{imp}}}\langle\boldsymbol{R}_{n}+\boldsymbol{r}_{i},\uparrow|G\left(\omega\right)|\boldsymbol{R}_{n}+\boldsymbol{r}_{i},\downarrow\rangle.
\end{eqnarray*}
The notation has been slightly changed to reflect the vector addition
of position vectors. While before only the site index was specified,
now the position $\boldsymbol{R}_{n}+\boldsymbol{r}_{i}$ is specified.
This matrix element can be expressed in terms of a random vector

\begin{equation}
|\xi,\sigma\rangle=\frac{1}{\sqrt{N_{\textrm{imp}}}}\sum_{n=1}^{N_{\textrm{imp}}}\xi_{\boldsymbol{R}_{n}}|\boldsymbol{R}_{n}+\boldsymbol{r}_{i},\sigma\rangle,\label{eq: random vector}
\end{equation}
thus casting the expression for $\rho_{0,i}$ into

\[
\rho_{0,i}=\imath\int_{-\Lambda}^{\Lambda}\text{d}\omega\left(1-2f\left(\omega\right)\right)\overline{\langle\xi,\uparrow|G\left(\omega\right)|\xi,\downarrow\rangle}
\]
where the bar denotes a random vector average. This expression can
now be calculated with direct Chebyshev expansion of the Green operator
in an efficient manner. 

\subsubsection{Order parameters \label{Appendix_F_SC - 2.Order-parameters}}

Using the mean-field model of the main text, we first calculated the
bulk superconducting order parameters $\Delta_{0}$ and $\Delta_{1}$
in the clean system. In this case, the parameters are homogeneous
and the sum in Eq. \ref{eq: random vector} runs over the whole lattice;
see Fig. \ref{fig_supp_05} (b) for $\Delta_{1}$. The calculation
was done at finite chemical potential $\mu=0.4t$. Both order parameters
vary continuously over the $g_{0},g_{1}$ plane. A qualitatively similar
picture exists for $\Delta_{0}$.

Then, we performed the same calculation, but with the disorder specified
in the main text (see Fig. \ref{fig_supp_05} a)). Now the order parameters
suffer a clear discontinuity which was not present before. The difference
between these two graphs is the result presented in the main text,
in a 2D color scheme {[}Fig. \ref{fig04}(c){]}.

\subsubsection{Renormalization\label{Appendix_F_SC - 3.Renormalization}}

The main effect of the impurities is to renormalize the energy scales
of the problem. Assuming the energy scales vary as $g_{0}^{\prime}=g_{0}\left[1+\alpha t/\left(t+\Delta t\right)\right]$
for some $\alpha$ depending on the concentration, we get the left
panel of Fig. \ref{fig_supp_05}(d). Here we used $\alpha=1.15$.
\end{document}